\documentclass[12pt]{article}
\usepackage{amsfonts}
\usepackage{amsmath}
\usepackage{amssymb}
\usepackage{amsthm}
\usepackage{color}
\usepackage{cite}
\usepackage{accents}
\usepackage{youngtab}
\usepackage{braket}
\usepackage{graphicx}
\usepackage[svgnames]{xcolor}
\usepackage{ytableau,tikz}
\usetikzlibrary{matrix}
\usepackage[svgnames]{xcolor}
\usepackage{comment}

\usepackage[colorlinks=true,linkcolor=black,citecolor=black,urlcolor=black]{hyperref}

\newcommand{\ba}{\begin{eqnarray}}
\newcommand{\ea}{\end{eqnarray}}
\newcommand{\nn}{\nonumber}

\newcommand{\g}{\mathfrak{g}}
\newcommand{\tn}{{\tilde n}}

\newcommand{\cU}{\mathcal{U}}
\newcommand{\cW}{\mathcal{W}}
\newcommand{\cE}{\mathcal{E}}

\newcommand{\CY}{\mathcal{Y}}
\newcommand{\lt}{\left}
\newcommand{\rt}{\right}
\newcommand{\CN}{\mathcal{N}}

\makeatletter

\@addtoreset{equation}{section}
\makeatother
\let\refOld\ref
\renewcommand{\ref}[1]{(\refOld{#1})}

 \DeclareMathOperator*{\Res}{Res}
 \DeclareMathOperator*{\tr}{tr} 
 \newcommand{\superp}[2]{\genfrac{}{}{0pt}{}{#1}{#2}}
 
 \def\d{\delta}
 
 \def\Re{{\rm Re ~}}
 
 \def\p{\partial}
 
 \def\a{\alpha}
 
 \def\g{\gamma}
 \def\d{\delta}
 \def\e{\epsilon}
 
 \def\th{\theta}
 
 \def\k{\kappa}
 \def\l{\lambda}
 
 \def\n{\nu}
 \def\x{\xi}

 \def\t{\tau}
 \def\th{\theta}
 \def\z{\zeta }

 \def\L{\Lambda}

\def\CE{{\mathcal{E}}}

\def\CN{{\mathcal{N}}}
\def\CO{{\mathcal{O}}}
\def\CP{{\mathcal{P}}}

\def\CU{{\mathcal{U}}}
\def\CV{{\mathcal{V}}}
\def\CW{{\mathcal{W}}}

\def\CY{{\mathcal{Y}}}
\def\CZ{{\mathcal{Z}}}
\def\la{\left\langle}
\def\ra{\right\rangle}

\def\hf{\dfrac{1}{2}}
\def\Op{\mathcal{O}}

\def\implies{\quad\Rightarrow\quad}

\def\tm{\tilde{\mu}}

\def\Zv{\mathcal{Z}_\text{vect.}}
\def\Zbf{\mathcal{Z}_\text{bfd.}}
\def\bZbf{\bar{\mathcal{Z}}_\text{bfd.}}
\def\Zf{\mathcal{Z}_\text{fund.}}
\def\Zaf{\mathcal{Z}_\text{a.f.}}
\def\Zinst{\mathcal{Z}_\text{inst.}}
\def\ZCS{\mathcal{Z}_\text{CS}}
\def\Uf{U_\text{fund.}}
\def\Uaf{U_\text{a.f.}}
\def\UCS{U_\text{CS}}

\def\res{\mathop{\text{Res}}}

\def\aY{|\vec{t},\vec{Y}\rangle}
\def\baY{\langle\vec{t},\vec{Y}|}
\def\Ga{|G,\vec{t}\rangle}
\def\bGa{\langle G,\vec{t}|}

\def\qf{\mathfrak{q}}
\def\PY{\Psi_{\vec Y}}
\def\pr{\mathrm{P}}

\def\tm{\tilde{m}}
\def\tn{\tilde{n}}
\def\pf{p_\text{fund.}}
\def\paf{p_\text{a.f.}}

\def\keff{\kappa_{eff}}
\def\tnf{{\tilde{n}_f}}

\begin{document}

\begin{titlepage}
	\renewcommand{\thefootnote}{\fnsymbol{footnote}}
	\vspace*{-2cm}
	\begin{flushright}
		UT-16-13
	\end{flushright}
	
	\vspace*{1cm}
		\begin{center}
{\bf {\LARGE \scalebox{1.15}{Coherent states in quantum $\cW_{1+\infty}$ algebra}}}\\
{\bf {\LARGE \scalebox{1.15}{and qq-character for 5d Super Yang-Mills}}}
		\end{center}
	\vspace{0.7cm}

	\begin{center}
		{\Large J.-E. Bourgine$^\dagger$, M. Fukuda$^\ast$, Y. Matsuo$^\ast$, H. Zhang$^\diamond$, R.-D. Zhu$^\ast$}
		\\[.4cm]
		{\em {}$^\dagger$INFN Bologna, Universit\`a di Bologna}\\
		{\em Via Irnerio 46, 40126 Bologna, Italy}
		\\[.4cm]
		{\em {}$^\ast$ Department of Physics, The University of Tokyo}\\
		{\em Bunkyo-ku, Tokyo, Japan}
		\\[.4cm]
		{\em {}$^\diamond$ Department of Physics and Center for Quantum Spacetime (CQUeST)}\\
		{\em Sogang University, Seoul 04107, Korea}
	\end{center}
	
	\vspace{0.7cm}
	
	\begin{abstract}
		\noindent
		
The instanton partition functions of $\mathcal{N}=1$ 5d super Yang-Mills are built using elements of the representation theory of quantum $\cW_{1+\infty}$ algebra: Gaiotto state, intertwiner, vertex operator. This algebra is also known under the names of Ding-Iohara-Miki and quantum toroidal $\widehat{\mathfrak{gl}}(1)$ algebra. Exploiting the explicit action of the algebra on the partition function, we prove the regularity of the 5d qq-characters. These characters provide a solution to the Schwinger-Dyson equations, and they can also be interpreted as a quantum version of the Seiberg-Witten curve.

\end{abstract}

	\vfill
	
\end{titlepage}
\vfil\eject

\setcounter{footnote}{0}

\section{Introduction}
Since the seminal work by Seiberg and Witten \cite{SW}, the relation between the BPS sector of $\CN=2$ supersymmetric Yang-Mills (SYM) theories in the Euclidean spacetime $\mathbb{R}^4$, and classical integrable models has attracted intensive studies (see \cite{Marshakov1999} and references therein). In 2004, Nekrasov \cite{Nekrasov2004} derived by localization the exact form of the $\CN=2$ SYM partition functions in a supergravity background, the Omega-background $\mathbb{R}_{\e_1}^2\times\mathbb{R}_{\e_2}^2$, depending on two deformation parameters $\e_1$ and $\e_2$. The exact partition function, including instanton corrections, was obtained as a perturbative sum over instanton sectors of coupled contour integrals. The evaluation of these integrals by Cauchy theorem produces a sum over residues in one-to-one correspondence with the set of boxes of Young diagrams. Sending the infrared cut-offs $\e_1,\e_2$ to zero, the Omega background reduces to $\mathbb{R}^4$. In this limit, the sum is dominated by large Young diagrams and the partition function reproduces the exponential of the prepotential obtained by Seiberg and Witten \cite{Nekrasov2003a}. On the other hand, in the limit $\e_2\to0$ while $\e_1$ fixed, the theory becomes effectively bi-dimensional and the BPS sector of $\CN=2$ SYM manifests the presence of quantum integrability \cite{Nekrasov2009,Nekrasov:2009b,Nekrasov2009c} in the form of a TBA-like non-linear integral equation \cite{Nekrasov2009,Meneghelli2013,Bourgine2014} and a set of Bethe equations \cite{Poghossian2010,Fucito2011,Fucito2012,Bourgine2014a,Nekrasov2012,NPS}.

In 2009, Alday, Gaiotto and Tachikawa \cite{Alday2010} conjectured a duality between Nekrasov partition functions of $\CN=2$ SYM with $U(2)$ gauge groups and 2d Liouville theory. This conjecture was then extended to $U(n)$ gauge groups and Toda theories in \cite{Wyllard2009,Mironov:2009by}. In several proofs of this conjecture \cite{Alba2011, SHc} and related , it has been essential to understand the action of the symmetry algebras of 2d conformal field theories (Virasoro for Liouville, $\CW_n$ for Toda) on a natural basis for the expansion of Nekrasov partition functions. This basis, called after Alba, Fateev, Litvinov and Tarnopolski (AFLT), is composed of states parameterized by $n$-tuple Young diagrams, where $n$ is the rank of the gauge groups \cite{Alba2011}. This representation is called \textit{rank $n$ representation}. The underlying symmetry of BPS $\CN=2$ SYM was found to be a deformation of $\cW_{1+\infty}$ which contains Virasoro and $\CW_n$ algebras combined with a $U(1)$ Heisenberg factor \cite{KMZ,MO}. As shown explicitly in \cite{Tsymbaliuk}(see also \cite{Prochazka}), this algebra turns out to be equivalent to the affine Yangian $Y(\widehat{\mathfrak{gl}_1})$ and to a spherical version of Cherednik's double degenerate Hecke algebra called SH$^c$ and built by Vasserot and Schiffmann in \cite{SHc}. The rank $n$ representation of SH$^c$ is equivalent to $\CW_n$ algebra for arbitrary $n$, even
in the case of degenerate representations \cite{FHSSY,fukuda2015sh}.

The AGT correspondence can be lifted to five dimensional $\CN=1$ SYM theories compactified on a circle, in correspondence with conformal blocks of q-Virasoro and q-Toda theories \cite{Awata2009,Awata2010,Tan2013,Taki2014}. The algebraic structures involved in the correspondence are the q-deformed versions of the previous ones. In this paper, we focus on the quantum continuous $\mathfrak{gl}_\infty$ algebra \cite{FFJMM1,Feigin:2010qea}, denoted $\CE$, which is equivalent to the Ding-Iohara algebra \cite{DI} with two additional Serre relations. In the degenerate limit, it reduces the affine Yangian $Y(\widehat{\mathfrak{gl}_1})$ \cite{Tsymbaliuk} in a similar manner as quantum loop algebras degenerate into Drinfeld second realization \cite{Drinfeld-Yangian} of Yangians of semisimple Lie algebras \cite{GL-Drinfeld}. $\CE$ is also conjectured in \cite{FFJMM1} to be isomorphic to the spherical double affine Hecke algebra (sDAHA) built in \cite{SV}. More importantly in the context of the  six-dimensional origin of AGT correspondence \cite{Mironov:2016cyq,Mironov:2016yue}, this algebra $\CE$ is essentially equivalent to the Ding-Iohara-Miki (DIM) \cite{DI,miki2007} algebra, also called quantum toroidal $\mathfrak{gl}_1$ \cite{Feigin2015,Feigin2016} or elliptic Hall algebra \cite{Burban2012}.\footnote{Quantum toroidal algebras \cite{Ginzburg2015} are obtained as the quantization of two variables loop algebras, here $\mathfrak{gl}_1[x^{\pm1},y^{\pm1}]$.}

The covariance of the partition function under the algebra $\CE$ originates a set of Ward identities between the correlators of the theory. These identities are equivalent to the regularity property of a sort of resolvent, the qq-character introduced in \cite{NS14,Nekrasov:2015wsu}.\footnote{The term \textit{resolvent} here refers to an analogy with matrix models developed in \cite{Nekrasov:2015wsu}.} This object corresponds to a (double) deformation of the character of (affine) Lie algebras. It has been interpreted in \cite{Kimura-Pestun} as the trace of the transfer matrix of a TQ-system that generalizes the concept of quantum Seiberg-Witten curve developed in the limit $\e_2\to0$ \cite{Poghossian2010,Fucito2011,Fucito2012,NPS,Mironov2009b,Mironov2010a,Mironov2010,Bourgine2012a}.\footnote{Alternatively, it can be constructed as a line defect in the 5d gauge theories, which also provides a different proof for its regularity \cite{Kim:2016qqs}.} Some primitive forms of Ward identities were studied in a series of works \cite{Zhang:2011au,Kanno:2012hk,KMZ,Matsuo:2014rba}.
In \cite{BMZ}, some of the authors developed a direct method to derive the regularity of qq-characters by exploiting the covariance of the 4d partition functions building blocks under the algebra 
 of symmetries, in their case SH$^c$. The building blocks for the correlation functions are given by the Gaiotto state (which describes the vector multiplet), the flavor vertex operators (fundamental hyper multiplets), and intertwiner operators (bifundamental hyper multiplets). A trivalent vertex supplements this construction in the case of gauge theories with non-linear quiver diagram. These few elements permit the construction of arbitrary $\CN=2$ quiver gauge theories with a $U(n)$ (or A-type) gauge group on each node. The action of the symmetry generators on these elements was expressed in terms of a single operator $\CY$ diagonal in the AFLT basis. The qq-character has a simple expression in terms of this operator, and its regularity followed from the transformation properties of the partition functions building blocks.

The aim of this paper is to provide a direct generalization of the construction exposed in \cite{BMZ} to quantum $\cW_{1+\infty}$. To do so, the transformation properties of the 5d building blocks under the generators of the algebra $\CE$ will be worked out. The regularity of the qq-characters for 5d $\CN=1$ SYM will follow from an equivalence between left and right actions in the expectation value of operators. The definition of the algebra and its representation are given in section two and three respectively. Section four presents the different building blocks of the partition functions, and section five provides the action of the algebra on these blocks. The regularity of the qq-characters is discussed in section six, it is interpreted as a quantization of Seiberg-Witten theory in section seven. It is shown in section eight that the results obtained in \cite{BMZ} are recovered in the limit of small radius of the compact dimension for which $\CE$ degenerates to SH$^
 c$. Several additional comments are gathered in the discussion section. Finally, the most relevant details of the calculations are presented in the appendix.

\section{Quantum $\cW_{1+\infty}$}\label{s:q-algebra}
The algebra $\CE$ depends on three Kerov deformation parameters $q_{\a}$ with $\a=1,2,3$ constraint to obey the relation $q_1q_2q_3=1$. 
The two independent parameters $q_1$ and $q_2$ are the K-theoretic versions of the 4d Omega background equivariant deformation parameters $\e_1$ and $\e_2$. Their dependencies will be encoded in the scattering function $h(z)$ and the parameter $\gamma_1$ defined as\footnote{In \cite{FFJMM1}, a function of two variables $g(z,w)$ is used instead of $h(z)$, they are related as
\begin{equation}
g(z,w)=\prod_{\a=1,2,3}(z-q_\a w),\quad h(z/w)=-\dfrac{g(z,w)}{g(w,z)},\nn
\end{equation}
with the implicit understanding in (\ref{e-e}) and (\ref{p-e}) that $z\neq q_\a^{\pm1} w$.}
\begin{equation}
h(z)=\prod_{\a=1,2,3}\dfrac{1-q_\a^{-1} z}{1-q_\a z},\quad \gamma_1=\prod_{\a=1,2,3}(1-q_\a).
\end{equation}
Note that the scattering function $h(z)$ obeys the unitarity property $h(z)h(z^{-1})=1$. The algebra $\CE$ is spanned by the set of generators $e_k$, $f_k$ and the Cartan elements $\psi_{k\geq0}^+$, $\psi_{k\leq 0}^-$. These generators form the Drinfeld currents
\ba
e(z)=\sum_{k\in\mathbb{Z}}e_kz^{-k}\,,\quad f(z)=\sum_{k\in\mathbb{Z}}f_kz^{-k}\,,\quad \psi^\pm (z)=\sum_{k\geq 0}\psi^{\pm}_{\pm k}z^{\mp k}\,,
\ea
and the algebra ${\cal E}$ is defined by the following set of relations:
\ba
e(z)e(w)=h(w/z)e(w)e(z)\,,\quad f(z)f(w)=h(z/w)f(w)f(z)\,,\label{e-e}\\
\psi^\pm (z)e(w)=h(w/z)e(w)\psi^\pm(z)\,,\quad \psi^\pm(z)f(w)=h(z/w)f(w)\psi^\pm(z)\,,\label{p-e}
\ea
\ba
\left[e(z),f(w)\right]=\frac1{\gamma_1}\delta\lt(\frac{z}{w}\rt)(\psi^+(z)-\psi^-(z))\,,\label{e-f}\\
\left[\psi^\pm(z),\psi^\pm(w)\right]=\left[\psi^+(z),\psi^-(w)\right]=0\,,\label{p-p}\\
\left[e_0,\left[e_1,e_{-1}\right]\right]=0\,,\quad \left[f_0,\left[f_1,f_{-1}\right]\right]=0\,,\label{definition-E}
\ea
with the $\delta$-function 
\ba
\delta(z)=\sum_{k\in\mathbb{Z}}z^k,\quad \d(z)F(z)=\d(z)F(1).
\ea
In addition, $\psi^\pm_0$ are central and invertible elements of the algebra. It is important to note that, in contrast with the works presented in \cite{Kimura-Pestun,Mironov:2016cyq,Mironov:2016yue,Awata:2016riz}, the algebra is considered here without the extra central element, or twist parameter. This difference turns out to have deep consequences on the representations which we will comment in the discussion section.


For the description of 5d SYM, the set of generators defined in $\CE$ is
not so convenient. Instead of using $\psi^\pm$, it is more useful to introduce the current
$D(z)=\sum_{k\in \mathbb{Z}} D_k z^{-k}$ which satisfies the following commutation relations with the Drinfeld currents,
\begin{equation}
[D(z),e(w)]=\d(z/w)e(z),\quad [D(z),f(w)]=-\d(z/w)f(w),\quad [D(z),D(w)]=0.
\end{equation}
These relations, together with (\ref{e-e},\ref{e-f}), provide an alternative definition of
the algebra.\footnote{In \cite{Feigin2016}, the operators $D_k$ are also introduced
in the algebra $\mathcal{E}$, and this extension is denoted $\mathcal{E}'$. For simplicity, here we keep the same name for the two possible presentations of the algebra.}
As we see below, the operators $\psi^\pm(z)$ are expressible in terms of
vertex operators built from the $D_k$ generators.  So in this second picture, $e(z), f(z), D(z)$
are the fundamental generators while $\psi^\pm(z)$ are the derived generators.

As in \cite{BMZ}, it is useful to introduce ``free boson" generators by a formal integration of the current $D(z)$, which will enable us to define later the ``vertex operators". It is useful to distinguish between the action of positive/negative modes, 
and define two operators as
\begin{equation}
\Phi_+(z)=\log(z)D_0-\sum_{k> 0}\dfrac{z^{-k}}{k}D_{k}\,,\quad \Phi_-(z)=-\sum_{k> 0}\dfrac{z^{k}}{k}D_{-k}\,.
\end{equation}
Using the corresponding vertex operators, we define the two $\CY$-operators $\CY^{\pm}(z)$ referred to as \textit{chiral ring generating operators} by Nekrasov, Pestun and Shatashvili (NPS) \cite{NPS}.\footnote{Our definition slightly differs from the one given in \cite{NPS} $\CY^\pm(z)=\CY_\text{NPS}^\pm(q_3^{\pm1/2}z)$.} These operators will play an essential role in the definition of qq-characters.
\begin{align}\label{Y-Phi}
\begin{split}
&\CY^+(z):=
e^{c^+(z)}\exp\left(\Phi_+(q_1^{-1}z)+\Phi_+(q_2^{-1}z)-\Phi_+(q_3z)-\Phi_+(z)\right),\\
&\CY^-(z):=
e^{c^-(z)}\exp\left(\Phi_-(q_1z)+\Phi_-(q_2z)-\Phi_-(q_3^{-1}z)-\Phi_-(z)\right),
\end{split}
\end{align}
where the series $c^\pm(z)$ are defined in terms of the central parameters of the algebra. In fact, the rank $n$ representation introduced in the next section depends on $n$ \textit{Coulomb branch parameters} $t_\ell$ that play the role of the representation weights. In term of these parameters, the central series are given by $c^+(z)=-\sum_{\ell=1}^n\sum_{k>0}\frac{t_{\ell}^k}{k}z^{-k}$ and $c^-(z)=-\sum_{\ell=1}^n\sum_{k>0}\frac{q_3^{-k}t_{\ell}^{-k}}{k}z^k$. As a result, the operators $\psi^\pm(z)$ can be expressed as, 
\begin{align}\label{psi_YY}
\begin{split}
\psi^+(z)
=(1-q_1^{-1})(1-q_2^{-1})\nu\CY^+(zq_3^{-1})\CY^+(z)^{-1},\\
\psi^-(z)
=(1-q_1^{-1})(1-q_2^{-1})q_3^n\nu\CY^-(z)\CY^-(zq_3)^{-1},
\end{split}
\end{align}
with
\begin{equation}
\nu=\prod_{\ell=1}^n\left(\dfrac{-1}{q_3t_\ell}\right).
\end{equation}

For the description of super Yang-Mills in 5d, we need to introduce an extra generator with the following commutation relations with $e(z)$ and $f(z)$,
\begin{eqnarray}\label{cU}
\cU e(z) =z e(z) \cU,\quad \cU f(z) =z^{-1}f(z)\cU\,.
\end{eqnarray}

Finally, it is worthwhile to introduce the decomposition of Drinfeld currents into positive and negative modes:
\begin{equation}
	e(z)=e_+(z)+e_-(z),\quad e_+(z)=\sum_{k=1}^\infty e_k z^{-k},\quad e_-(z)=\sum_{k=0}^\infty e_{-k}z^k,
\end{equation}
and similarly for $f(z)=f_+(z)+f_-(z)$. With this decomposition, the commutation relation (\ref{e-f}) takes the following form:
\begin{equation}\label{com_ef}
[e_\eta(z),f_{\eta'}(w)]=-\dfrac{\eta\eta'}{\g_1}\dfrac{z\psi^\eta(z)-w\psi^{\eta'}(w)}{z-w}+\dfrac{\eta\eta'}{\g_1}\psi_0^+,
\end{equation}
where $\eta,\eta'=\pm$, and it has been assumed $|z|<|w|$ for $\eta=\eta'=+$ and $|w|<|z|$ for $\eta=\eta'=-$. The derivation of this identity is presented in appendix \ref{AppA}.

By definition, the algebra is clearly invariant under the permutations of $q_1$, $q_2$ and $q_3$. In the degenerate version of this algebra known as SH$^c$, this $\mathfrak{S}_3$ transformation was referred to as a triality automorphism \cite{fukuda2015sh} which is related to
the level-rank duality in WZW coset models \cite{kuniba1991ferro,Altschuler:1990th}
for finite rank representations.

The algebra $\CE$ also exhibits an invariance under two discrete symmetries involving the inversion of the parameters $q_\a$. When these parameters $q_\a$ are substituted by their inverse, the scattering function $h(z)$ is replaced by its inverse and the sign in front of $\g_1$ is flipped. There are two ways to render the algebra invariant under this symmetry. The simplest one is to exchange the two Drinfeld currents $e(z)\leftrightarrow f(z)$. A more involved realization consists in exchanging the positive and negative modes of the generators, $e_k\leftrightarrow e_{-k}$, $f_k\leftrightarrow f_{-k}$, $\psi^+_k\leftrightarrow\psi^-_k$,
while also inversing the spectral parameter $z\leftrightarrow z^{-1}$. This automorphism can be physically interpreted as a parity transformation along the $S^1$ compact direction.

\section{Rank $n$ representation}\label{s:FT-rep}
The rank $n$ representation is obtained as an $n$-tensor product of the action over Macdonald polynomials \cite{FHHSY} using the coproduct of $\CE$ \cite{FHSSY,Feigin-Tsymbaliuk,Tsymbaliuk}. Thus, the Hilbert space $\CV$ is spanned by the AFLT states $\aY$ parameterized by $n$ Young diagrams $\vec Y=(Y_\ell)_{\ell=1}^n$. The $n$-vector $\vec t$ defines a set of central charges (or weights) characterizing the space of representation. This basis is orthonormal, and is sometimes called the ``fixed-point basis'' \cite{Tsymbaliuk}, because each state corresponds to a fixed point in the calculation of Nekrasov's partition function for A-type gauge theories. The shape of Young diagrams can be encoded in one of the two sets $A(\vec Y)$ and $R(\vec Y)$ respectively corresponding to the set of boxes that can be added to or removed from the diagrams. In addition, we introduce for each box $x=(\ell,i,j)$ with $(i,j)\in Y_\ell$ a coordinate-like number $\chi_x=t_\ell q_1^{i-1}q_2^{j-1}$.

\subsection{Representation of algebra $\cal{E}$}
The action of the generators  $e(z)$, $f(z)$, $\psi^\pm(z)$ on the AFLT basis
may be found in \cite{Feigin-Tsymbaliuk,Tsymbaliuk}.
We make a minor modification of the action of $e(z)$ and $f(z)$
and change the normalization of the basis\footnote{The precise form of the modification of the basis and generators is
given in appendix \ref{a:algebra}.}
to render the coefficients symmetric between $e(z)$ and $f(z)$, up to the overall factor $z^{-n}$ as in \cite{KMZ}.
\begin{eqnarray}
e(z)\ket{\vec{t},\vec{Y}}&=&\sum_{x\in A(\vec{Y})}\delta(z/\chi_x)\Lambda_{x}(\vec{Y})\ket{\vec{t},\vec{Y}+x}\,,\label{eket}\\
f(z)\ket{\vec{t},\vec{Y}}&=&z^{-n}\sum_{x\in R(\vec{Y})}\delta(z/\chi_x)\Lambda_{x}(\vec{Y})\ket{\vec{t},\vec{Y}-x}\,,\label{fket}\\
\psi^{\pm}(z)\ket{\vec t,\vec Y}&=&\left[\Psi_{\vec Y}(z)\right]_\pm\ket{\vec t,\vec Y}\,.
\label{pket}
\end{eqnarray}
Here $\vec Y\pm x$ denotes the addition/subtraction 
of a box $x$ from the $n$-tuple Young diagram $\vec Y$.  
The coefficients $\Psi_{\vec Y}(z)$ and $\Lambda_x(\vec Y)$ are defined by
\begin{eqnarray}
\Psi_{\vec Y}(z)&=&(1-q_1^{-1})(1-q_2^{-1})
\nu \prod_{x\in A(\vec Y)}\frac{z-q_3\chi_x}{z-\chi_x}
\prod_{x\in R(\vec Y)}\frac{z-q_3^{-1}\chi_x}{z-\chi_x},\\
\L_x(\vec Y)^2&=&\mp\dfrac1{\g_1}\chi_x^{n-1}\Res_{z\to \chi_x}\Psi_{\vec Y}(z)
=\prod_{\substack{y\in A(\vec{Y})\\ y\neq x}}
\frac{1-\chi_x\chi_y^{-1} q_3^{-1}}{1-\chi_y \chi_x^{-1} }
\prod_{\substack{y\in R(\vec{Y})\\ y\neq x}}\frac{1-\chi_y\chi_x^{-1} q_3^{-1}}{1-\chi_x\chi_y^{-1}}\,.\label{Lx}
\end{eqnarray}
The sign in the second term in (\ref{Lx}) 
is negative for $x\in A(\vec Y)$ and positive when $x\in R(\vec Y)$. 
The consistency of the representation 
(\ref{eket}--\ref{pket}) with the algebra (\ref{e-e}--\ref{p-p})
is checked in appendix \ref{a:algebra}. The eigenvalues $\Psi_{\vec Y}(z)$ are rational functions with simple poles at $z=\chi_x$ for $x\in A(\vec Y)\cup R(\vec Y)$, and their residues coincide with the coefficients in the action of $e(z), f(z)$.

We added brackets $[\cdots]_\pm$ in (\ref{pket}).
Even though the eigenvalue $\Psi_{\vec Y}(z)$ is common 
to both operators $\psi^\pm(z)$, it should be expanded 
in powers of $z^{-1}$ if we consider the action of $\psi^+$, 
while in powers of $z$ for the action of $\psi^-$. 
$[\cdots]_\pm$ implies that the expression inside the brackets should be expanded
in terms of $z^{\mp 1}$.
Such a distinction is essential, and may be better understood in the following example.
Let us consider the two expansions of
\begin{equation}\label{example}
\frac{z}{z-w}=\left\{\begin{array}{ll} \sum_{n=0}^\infty (w/z)^n \quad & |z|>|w|\\
- \sum_{n=1}^\infty (z/w)^n \quad & |z|<|w|\end{array}
\right.
\end{equation}
While the left hand side of both series is formally identical, the difference between the
right hand side is non-vanishing $\sum_{n\in \mathbb{Z}}(z/w)^n=\delta(z/w)$.
Likewise, the difference between the actions of $\psi^{\pm}(z)$ does not vanish, but is instead represented by a sum of $\d$-functions centered at the poles of $\Psi_{\vec Y}(z)$,
\begin{eqnarray}
(\psi^{+}(z)-\psi^{-}(z))\ket{\vec t,\vec Y}&=&
\gamma_1(-\sum_{x\in A(\vec Y)}+\sum_{x\in R(\vec Y)})
\chi_x^{-n} \delta(z/\chi_x)\Lambda_x(\vec Y)^2 
\ket{\vec t,\vec Y}\,.
\end{eqnarray}
In the following, we meet some operators with $\pm$ index which have apparently
the same eigenvalue.  The notation $[\cdots]_\pm$ will be used to make
the distinction explicit.

The action of positive and negative modes of $e(z), f(z)$ is easily deduced from (\ref{eket},\ref{fket}):
\begin{align}\label{repr_ef}
\begin{split}
&e_+(z)
\ket{\vec{t},\vec{Y}}=-\sum_{x\in A(\vec Y)}\left[\dfrac{\L_x(\vec Y)}{1-z\chi_x^{-1}}\right]_+\ket{\vec{t},\vec{Y}+x},\quad 
f_{+}(z)
\ket{\vec{t},\vec{Y}}=-\sum_{x\in R(\vec Y)}\left[\dfrac{\L_x(\vec Y)\chi_x^{-n}}{1-z\chi_x^{-1}}\right]_+\ket{\vec{t},\vec{Y}-x},\\
&e_-(z)\ket{\vec{t},\vec{Y}}=\sum_{x\in A(\vec Y)}\left[\dfrac{\L_x(\vec Y)}{1-z\chi_x^{-1}}\right]_-\ket{\vec{t},\vec{Y}+x},\quad
f_-(z)\ket{\vec{t},\vec{Y}}=\sum_{x\in R(\vec Y)}\left[\dfrac{\L_x(\vec Y)\chi_x^{-n}}{1-z\chi_x^{-1}}\right]_-\ket{\vec{t},\vec{Y}-x}.\\
\end{split}
\end{align}

We define the bra basis by $\langle\vec t, \vec Y|\vec t, \vec Y'\rangle=\delta_{\vec Y,\vec Y'}$.  The action of any generator $\mathcal{O}$
on the bra basis is determined by $(\langle \vec t, \vec Y| \mathcal{O})|\vec t, \vec Y'\rangle=
\langle \vec t, \vec Y| (\mathcal{O}|\vec t, \vec Y'\rangle$.
In particular, we have,
\begin{eqnarray}
\langle \vec t, \vec Y|e(z)&=& \sum_{x\in R(\vec Y)}
\delta(z/\chi_x) \Lambda_x(\vec Y)\langle \vec t,\vec Y-x|\\
\langle \vec t, \vec Y|f(z)&=& z^{-n}\sum_{x\in A(\vec Y)}
\delta(z/\chi_x) \Lambda_x(\vec Y)\langle \vec t,\vec Y+x|\,.
\end{eqnarray}
In the derivation of these formulae, we have used the identities,
$\Lambda_x(\vec Y-x)=\Lambda_x(\vec Y)$ for $x\in R(\vec Y)$
and $\Lambda_x(\vec Y+x)=\Lambda_x(\vec Y)$ for $x\in A(\vec Y)$.

\subsection{Eigenvalues for the extra Cartan generators and vertex operators}
As emphasized in the previous section, it is more useful to extend
the algebra $\mathcal{E}$ by the introduction of the extra diagonal generators $D(z),\mathcal{U}$, together with the vertex operator written in terms of $\Phi_{\pm}$. The action of these generators on the basis is written in a compact diagonal form. For instance, the action of $D(z)$ on the AFLT basis reads
\begin{equation}\label{def_Dz}
D_k\aY=\sum_{x\in \vec Y}\left(\chi_x\right)^k\aY,\quad\text{or}\quad D(z)\aY=\sum_{x\in \vec Y}\d(z/\chi_x)\aY\quad\text{with}\quad D(z)=\sum_{k\in\mathbb{Z}}D_kz^{-k}.
\end{equation}

The action of exponentiated operators on the AFLT basis
can be derived from (\ref{def_Dz}).  The expansions at $z=\infty$ (for $\Phi_+$) or $z=0$
can be re-summed and produce simple products:
\begin{equation}\label{def_Phi}
e^{\Phi_+(z)}
\aY
=\left[\prod_{x\in\vec Y}(z-\chi_x)\right]_+ \aY,\quad 
e^{\Phi_-(z)}
\aY=\left[\prod_{x\in\vec Y}(1-z\chi_x^{-1})\right]_-\aY.
\end{equation}
Since the expressions in the RHS have no singularities, their asymptotic series expansion are convergent in the whole complex plane, and the vertex operators can be safely analytically continued.

Using a specialization of the \textit{shell formula} which describes the cancellation between the factors associated to the boxes of a Young diagram to produce only edge contributions,
\begin{equation}\label{shell}
\prod_{\ell=1}^n(z-t_\ell)\prod_{x\in \vec Y}\dfrac{(z-q_1\chi_x)(z-q_2\chi_x)}{(z-\chi_x)(z-q_3^{-1}\chi_x)}=\dfrac{\prod_{x\in A(\vec Y)}(z-\chi_x)}{\prod_{x\in R(\vec Y)}(z-\chi_xq_3^{-1})},
\end{equation}
it is possible to evaluate the action of  $\CY^\pm(z)$ on the AFLT basis,
\begin{equation}
\CY^\pm(z)\ket{\vec{t},\vec{Y}}=\left[\tilde\CY^\pm(z,\vec Y)\right]_\pm \ket{\vec{t},\vec{Y}},
\end{equation}
with the eigenvalues expanded at $z=\infty$ and $z=0$ respectively,
\begin{equation}
\label{def_Y}
\tilde\CY^+(z,\vec Y)=
\frac{\prod_{x\in A(\vec{Y})}1-z^{-1}\chi_x}{\prod_{x\in R(\vec{Y})}1-z^{-1}\chi_xq_3^{-1}} \,,\quad
\tilde\CY^-(z,\vec Y)=\frac{\prod_{x\in A(\vec{Y})}1-z\chi_x^{-1}q_3^{-1}}{\prod_{x\in R(\vec{Y})}1-z\chi_x^{-1}}\,.
\end{equation}
It can be shown using the shell formula that these rational functions are related through the formula
\begin{equation}\label{rel_Y}
\tilde \CY^-(zq_3,\vec Y) = \nu (zq_3)^n \tilde\CY^+(z,\vec Y).
\end{equation}
Finally the action of the operator $\mathcal{U}$ on the AFLT basis is written
\begin{eqnarray}
\cU \aY = \prod_{x\in \vec Y} \chi_x \aY\,,
\end{eqnarray}
it will be employed to describe the contribution of Chern-Simons terms to the partition function of 5d super Yang-Mills.

\section{Nekrasov partition function and discrete Ward identities}
\subsection{Building blocks}
As in the description of the 4d case in \cite{BMZ}, 5d Nekrasov instanton partition functions can be obtained by combining specific coherent states and operators of the finite rank representations of $\CE$. They are the $q$-deformed version of the Gaiotto state, intertwiner, dilatation and flavor vertex operators and trivalent vertex. In this section, we define these building blocks and explain how the partition function is constructed by combining them.

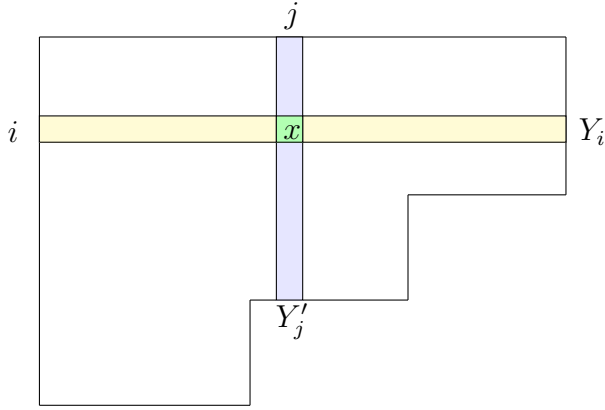
\begin{figure}[bpt]
\begin{center}
\begin{tikzpicture}[scale=0.7]
\draw (0,0) -- (10,0);
\draw (10,0) -- (10,-3);
\draw (10,-3) -- (7,-3);
\draw (7,-3) -- (7,-5);
\draw (7,-5) -- (4,-5);
\draw (4,-5) -- (4,-7);
\draw (4,-7) -- (0,-7);
\draw (0,0) -- (0,-7);
\filldraw [fill=yellow!20] (0,-1.5) rectangle (10,-2);
\filldraw [fill=blue!10] (4.5,0) rectangle (5,-5);
\filldraw [fill=green!30] (4.5,-1.5) rectangle (5,-2);
\draw (4.8,0.4) node {$j$};
\draw (-0.5,-1.8) node {$i$};
\draw (10.5,-1.8) node {$Y_i$};
\draw (4.8,-5.4) node {$Y'_j$};
\draw (4.8,-1.8) node {$x$};
\end{tikzpicture}
\end{center}
\caption{Convention for the labeling of a box $x=(i,j)$ in Young diagrams.}
\label{convention}
\end{figure}

\paragraph{Gaiotto state}
The simplest $\CN=1$ SYM theory contains only a single gauge multiplet with gauge group $U(n)$. It is associated with the rank $n$ representation space $\CV$ of the algebra $\CE$. In this case, the Nekrasov partition function\footnote{We regard here only the instanton contribution to the full partition function. The perturbative part, consisting of classical and one-loop contributions, will not be discussed.} is obtained as the norm of the Gaiotto state defined in \cite{Awata2009,Awata2010,Taki2014} as a Whittaker state for the q-Virasoro (or q-W) algebra.  The construction for the algebra $\CE$ is given in \cite{Awata2011,feigin2011}. This q-deformed version of the Virasoro Gaiotto state \cite{Gaiotto:2009ma,Marshakov:2009gn,Kanno:2011fw,Kanno:2012xt} produces the instanton partition function by inner product:
\ba\label{Zinst_Un}
\Zinst=\bra{G,\vec{t}}\mathfrak{q}^D\ket{G,\vec{t}}\ ,
\ea
where the grading by the exponentiated gauge coupling $\qf=\exp(2\pi i\tau)$ has been factorized out of the states. The dilatation operator $D=D_0$ corresponds to the zero mode of the operator $D(z)$ defined in (\ref{def_Dz}), it is equivalent to the Virasoro zero-mode
$L_0$. This operator is diagonal in the AFLT basis $\aY$ with eigenvalue $|\vec Y|$ equals to the total number of boxes in the $n$-tuple diagram $\vec Y$. The Gaiotto state corresponds to a sum over all the states $\aY$ with a weight expressing the vector multiplet contribution to the partition function,
\ba
\ket{G,\vec{t}}:=\sum_{\vec{Y}}\sqrt{\Zv(\vec{t},\vec{Y})}\ket{\vec{t},\vec{Y}}\,,
\ea
where the vector contribution is a product over Nekrasov factors $N_{Y_k,Y_l}(z)$ indexed by two Young diagrams \cite{Nakajima-Yoshioka}:
\ba\label{def_Zv}
\Zv(\vec{t},\vec{Y})^{-1}&:=&\prod_{k,l=1}^n N_{Y_k,Y_l}(t_k t_l^{-1})\\
N_{Y_k,Y_l}(t)&=&\prod_{(i,j)\in Y_k}
\lt(1-t q_1^{-Y^{(l)\prime}_j+i}
q_2^{Y_i^{(k)}-j+1}\right)
\prod_{(i,j)\in Y_l}\left(1-t q_1^{Y^{(k)\prime}_j-i+1}q_2^{-Y_i^{(l)}+j}\right)\,.\label{Nekrasov-factor}
\ea
In this expression, $Y'$ denotes the transposed of the Young diagram $Y$, and $Y_i$ is the number of boxes in the $i$-th row of the diagram $Y$.\footnote{Convention for box labels inside Young diagrams has been reversed from what was used in \cite{BMZ}: $(i,j)$ now denotes the box in the $i$-th row and $j$-th column (see Figure \ref{convention}).} The Nekrasov partition function for pure $U(n)$ gauge theory is reproduced from (\ref{Zinst_Un}) using the orthonormality of the AFLT states,
\ba
\Zinst=\sum_{\vec{Y}}\mathfrak{q}^{|\vec{Y}|}\Zv(\vec{t},\vec{Y})\,.
\ea

\paragraph{Intertwiner}
When the quiver diagram of the gauge theory has more than one node, we need to introduce several representation spaces, one for each node. The interwiner inserted between nodes $k$ and $l$ is defined as an operator acting in the space $\CV_l$, and taking value in the space $\CV_k$,\footnote{This intertwiner is affiliated to the vertex operator constructed in \cite{Awata2011} upon the Ding-Iohara algebra. For the rank two representation, it is expected to be equivalent to the q-Virasoro vertex operator constructed in \cite{Itoyama:2016dwe} (see also \cite{Awata:1996xt,Kadeishvili:1996ik,Jimbo1997} for previous works).}
\begin{equation}
V_{kl}(\vec{t}_k,\vec{t}_l|\mu_{kl}):\CV_l\to\CV_k
\end{equation}
In addition to the central parameters $\vec t_k$ and $\vec t_l$ of the two representation spaces, it also depends on an extra twist parameter $\mu_{kl}$. In the gauge theory, the intertwiner describes the contribution of bifundamental fields transforming under the gauge group $U(n_k)\times U(n_l)$, and $\mu_{kl}$ denotes the mass of these fields. Hence, the definition of the intertwiner involves the bifundamental contribution to the instanton partition function, suitably normalized by the vector contributions of the two gauge groups:
\begin{align}
\begin{split}\label{def_intertwn}
&V_{k,l}(\vec{t}_k,\vec{t}_l|\mu_{kl})=\sum_{\vec{Y}_k,\vec{Y}_l}\bZbf(\vec{t}_k,\vec{Y}_k;\vec{t}_l,\vec{Y}_l|
\mu_{kl})\ket{\vec{t}_k,\vec{Y}_k}\bra{\vec{t}_l,\vec{Y}_l}\,,\\
&\bZbf(\vec{t}_k,\vec{Y}_k;\vec{t}_l,\vec{Y}_l|\mu_{kl})=\sqrt{\Zv(\vec{t}_k,\vec{Y}_k)\Zv(\vec{t}_l,\vec{Y}_l)}
\Zbf(\vec{t}_k,\vec{Y}_k;\vec{t}_l,\vec{Y}_l|\mu_{kl})\,.
\end{split}
\end{align}
The bifundamental partition function is a well-known quantity that can be found for instance in \cite{Awata2008,Awata2009}. It can be expressed as a product over the Young diagrams composing the $n$-tuples $\vec Y$ and $n'$-tuple $\vec W$ of the Nekrasov factor (\ref{Nekrasov-factor}),
\ba\label{def_bif}
\Zbf(\vec{t},\vec{Y};\vec{t'},\vec{W}|\mu)=\prod_{p=1}^{n}\prod_{q=1}^{n'}N_{Y_p,W_q}(t_pt_q^{\prime-1}\mu^{-1})\,.
\ea
It is readily observed that the vector contribution given in (\ref{def_Zv}) is a particular case of the bifundamental contribution, obtained for two identical $n$-tuple Young diagrams and a bifundamental mass of unity:
\begin{equation}\label{Zv_Zbf}
\Zv(\vec{t},\vec{Y})=\Zbf(\vec{t},\vec{Y};\vec{t},\vec{Y}|1)^{-1}.
\end{equation}
Once defined the intertwiner, pure $\CN=1$ SYM instanton partition function with gauge group $U(n_1)\times U(n_2)$ and exponentiated gauge couplings $\qf_1$ and $\qf_2$ can be written as a double expectation value in each representation space,
\begin{equation}
\Zinst=\langle G,\vec t_1|\qf_1^D V_{12}(\vec t_1,\vec t_2|\mu) \qf_2^D |G, \vec t_2\rangle=\sum_{\vec Y_1,\vec Y_2}\qf_1^{|\vec Y_1|}\qf_2^{|\vec Y_2|}\Zv(\vec t_1,\vec Y_1)\Zv(\vec t_2,\vec Y_2)\Zbf(\vec t_1,\vec Y_1;\vec t_2,\vec Y_2|\mu).
\end{equation}
This expression generalizes without effort to linear quivers in the absence of fundamental/antifundamental matter fields. Affine $\hat A$ quiver partition functions are obtained by taking the trace over the representation space. For instance, for a single gauge group $U(n)$,
\begin{equation}
\Zinst=\tr_{\CV}\left[\qf^D V_{12}(\vec t;\vec t|\mu)\right]=\sum_{\vec Y}\qf^{|Y|}\Zv(\vec t,\vec Y)\Zbf(\vec t,\vec Y;\vec t;\vec Y|\mu)=\sum_{\vec Y}\qf^{|Y|}\Zv(\vec t,\vec Y)\mathcal{Z}_{\mathrm{adj.}}(\mu;\vec t,\vec Y),
\end{equation}
where the trace $\tr$ is defined as a sum over the vectors $\aY$ spanning the AFLT basis, with a weight one:
\begin{equation}
\tr_{\CV}\Op=\sum_{\vec Y}\baY\Op\aY.
\end{equation}

\paragraph{Chern-Simons term} In five dimensions, a Chern-Simons term with level $\k$ can be added to the action without spoiling the supersymmetry. As a result, the partition function associated with the fixed point 
of the localization is modified by the inclusion of the factor:
\begin{equation}
\ZCS(\k,\vec Y)=\prod_{x\in\vec Y}\left(\chi_x\right)^\k.
\end{equation}
As we previously mentioned, the inclusion of such a term is implemented by the insertion of the operator $(\cU)^\kappa$ in the inner product of the representation space associated to the corresponding node. The insertion of this operator is equivalent to a modification of the representation by a redefinition of the basis,
\begin{eqnarray}
\aY_\k :=(\cU)^\k \aY.
\end{eqnarray}
In the new representation space $\CV^{(\k)}$, the action of the Drinfeld currents is modified as follows,
\begin{eqnarray}
e(z)\aY_\k &=&z^{-\k}\sum_{x\in A(\vec{Y})}\delta(z/\chi_x)\Lambda_{x}(\vec{Y})\ket{\vec{t},\vec{Y}+x}_\k\,,\\
f(z)\ket{\vec{t},\vec{Y}}_\k&=&z^{\k-n} \sum_{x\in R(\vec{Y})}
\delta(z/\chi_x)\Lambda_{x}(\vec{Y})\ket{\vec{t},\vec{Y}-x}_k\,.
\end{eqnarray}
As before, the action on bra vectors is determined from the orthonormality condition 
\begin{equation}\label{def_scalar}
_{\kappa'}\langle\vec{t},\vec{Y}\ket{\vec{t},\vec{Y}'}_\kappa=\prod_{x\in \vec{Y}}(\chi_x)^{\kappa-\kappa'}\delta_{\vec{Y},\vec{Y}'}.
\end{equation}

\paragraph{Flavor vertex operators} To introduce matter fields in 5d SYM, it is necessary to distinguish between fundamental and antifundamental representations of the gauge groups, since the two contributions are different. These contributions depend on a mass vector $\vec m$ (resp. $\vec\tm$ for the antifundamental) with $n_f$ (resp. $\tn_f$) components, in addition to the parameters $\vec t$, $n$ characterizing the node with which the matter is coupled. They are obtained from the bifundamental contribution (\ref{def_bif}) by specialization to an empty $n_f$-tuple Young diagram,
\begin{equation}\label{def_Zf_Zaf}
\Zf(\vec m;\vec t,\vec Y)=\Zbf(\vec t,\vec Y;\vec m,\vec\emptyset|1),\quad \Zaf(\vec \tm;\vec t,\vec Y)=\Zbf(\vec\tm,\vec \emptyset;\vec t,\vec Y|1).
\end{equation}
Explicitly,
\begin{equation}
\Zf(\vec m;\vec t,\vec Y)=\prod_{f=1}^{n_f}\prod_{x\in\vec Y}(1-\chi_xq_3^{-1}m_f^{-1}),\quad \Zaf(\vec \tm;\vec t,\vec Y)=\prod_{f=1}^{\tn_f}\prod_{x\in\vec Y}(1-\tm_f\chi_x^{-1}).
\end{equation}
These factors can be obtained by acting on the states of the rank $n$ representation with the vertex operators defined in (\ref{def_Phi}),
\begin{eqnarray}
&&\Uf^+(\vec m)\aY=\left[\Zf(\vec m;\vec t,\vec Y)\right]_+\aY,\quad
\Uaf^-(\vec\tm)\aY=\left[\Zaf(\vec\tm;\vec t,\vec Y)\right]_-\aY,
\label{UZfnd}\\
&& \Uf^+(\vec m):=\prod_{f=1}^{n_f}(m_fq_3)^{-D}e^{\Phi_+(m_fq_3)},\quad \Uaf^-(\vec\tm):=\prod_{f=1}^{\tn_f}e^{\Phi_-(\tm_f)},
\end{eqnarray}
where the action of the zero mode of $\Phi_+(z)$ has been canceled using the dilatation operator $D$. These operators are diagonal in the AFLT basis and reproduce the contributions (\ref{def_Zf_Zaf}) when acted upon these sates. For instance, matter $\CN=1$ SYM with $U(n)$ gauge group is obtained as
\begin{equation}\label{Zinst_fund}
\Zinst=\bGa \qf^D\Uf^+(\vec m)\Uaf^-(\vec\tm)\Ga=\sum_{\vec Y}\qf^{|\vec Y|}\Zv(\vec t,\vec Y)\Zf(\vec m;\vec t,\vec Y)\Zaf(\vec \tm;\vec t,\vec Y).
\end{equation}
With the use of the Chern-Simons operator, one may obtain an alternative expression for $\Uf^+$ and $\Uaf^-$:
\begin{eqnarray}
\Uf^-(\vec m)&=& \CU^{n_f} \prod_{f=1}^{n_f} (-m_f q_3)^{-D} e^{\Phi_-(m_fq_3)}\,.\\
\Uaf^+(\vec\tm) &=& \CU^{-\tilde{n}_f}\prod_{f=1}^{\tn_f}(-1)^D e^{\Phi_+(\tilde{m}_f)}\,.
\end{eqnarray}
Finally in (\ref{UZfnd}) we have the symbol $\left[\cdots\right]_\pm$ 
to imply the expansion with respect to the
mass parameters $m_f, \tilde m_f$ at $0$ or $\infty$
since they appear as the arguments of $\Phi_\pm$. 
We note, however, that the two formal expansions $\left[\Zf\right]_\pm$
coincide after the summation since there are no simple poles in $\Zf, \Zaf$.  
In this sense, we will abbreviate the formal
expansion symbol $[\cdots]_\pm$ and treat the parameters $m_f,\tilde m_f$ as taking finite values
in the following.

\paragraph{Trivalent vertex}
For the description of the quiver gauge theory with bifurcation, we need to further introduce a trivalent vertex,
\ba
\ket{T,\vec t}_{k,l,m} = \sum_{\vec Y}
\Zv(\vec t, \vec Y)^{-1/2} \ket{\vec t, \vec Y}_k\otimes  \ket{\vec t, \vec Y}_l \otimes  \ket{\vec t, \vec Y}_m\,.
\ea
We refer to the discussion section of \cite{BMZ} for some
examples.  Unlike the other building blocks, the action of $\mathcal{E}$
generators on the trivalent vertex is difficult to evaluate at this moment.
For this reason, our discussion in this paper is limited to the linear (and affine) 
quiver gauge theories.

\subsection{Discrete Ward identities}
In appendix \ref{a:recur-Nekra}, a set of identities is derived by examination of the variation of the most general Nekrasov-type factors under the addition or subtraction of boxes in the Young diagrams. These identities are the analogue of the loop equations in Random Matrix Models, they will be called here \textit{discrete Ward identities} to emphasize the fact that they encode the covariance under the symmetry algebra acting on instanton partition functions. They are the q-analogue of the recursion relations studied in \cite{KMZ,BMZ}.

\ba
\frac{\Zbf(\vec{t},\vec{Y}+x;\vec{t'},\vec{W}|\mu)}{\Zbf(\vec{t},\vec{Y};\vec{t'},\vec{W}|\mu)}=\frac{\prod_{y\in A(\vec{W})}1-\chi_x\chi_y^{-1}q_3^{-1}\mu^{-1}}
{\prod_{y\in R(\vec{W})}1-\chi_x\chi_y^{-1}\mu^{-1}}\,,\label{recursion-s}\\
\frac{\Zbf(\vec{t},\vec{Y}-x;\vec{t'},\vec{W}|\mu)}{\Zbf(\vec{t},\vec{Y};\vec{t'},\vec{W}|\mu)}=\frac{\prod_{y\in R(\vec{W})}1-\chi_x\chi_y^{-1}\mu^{-1}}
{\prod_{y\in A(\vec{W})}1-\chi_x\chi_y^{-1}q_3^{-1}\mu^{-1}}\,,\\
\frac{\Zbf(\vec{t},\vec{Y};\vec{t'},\vec{W}+x|\mu)}{\Zbf(\vec{t},\vec{Y};\vec{t'},\vec{W}|\mu)}=\frac{\prod_{y\in A(\vec{Y})}1-\chi_y\chi_x^{-1}\mu^{-1}}
{\prod_{y\in R(\vec{Y})}1-\chi_y\chi_x^{-1}q_3^{-1}\mu^{-1}}\,,\\
\frac{\Zbf(\vec{t},\vec{Y};\vec{t'},\vec{W}-x|\mu)}{\Zbf(\vec{t},\vec{Y};\vec{t'},\vec{W}|\mu)}=\frac{\prod_{y\in R(\vec{Y})}1-\chi_y\chi_x^{-1}q_3^{-1}\mu^{-1}}
{\prod_{y\in A(\vec{Y})}1-\chi_y\chi_x^{-1}\mu^{-1}}\,.\label{recursion-e}
\ea
The recursion formulae for the vector multiplet can be obtained by using the relation (\ref{Zv_Zbf}) after a careful treatment of the contact terms in the limit $\mu\to1$:
\ba\label{discrete_Ward_Zv}
\frac{\Zv(\vec{t},\vec{Y}+x)}{\Zv(\vec{t},\vec{Y})}=\frac{1}{(1-q_1)(1-q_2)}\frac{\prod_{y\in R(\vec{Y})}(1-\chi_x\chi_y^{-1})(1-\chi_y\chi_x^{-1}q_3^{-1})}
{\prod_{\substack{y\in A(\vec{Y})\\y\neq x}}(1-\chi_x\chi_y^{-1}q_3^{-1})(1-\chi_y\chi_x^{-1})}\,,\\
\frac{\Zv(\vec{t},\vec{Y}-x)}{\Zv(\vec{t},\vec{Y})}=\frac{1}{(1-q_1)(1-q_2)}\frac{\prod_{y\in A(\vec{Y})}(1-\chi_x\chi_y^{-1}q_3^{-1})(1-\chi_y\chi_x^{-1})}
{\prod_{\substack{y\in R(\vec{Y})\\y\neq x}}(1-\chi_x\chi_y^{-1})(1-\chi_y\chi_x^{-1}q_3^{-1})}\,.
\ea

\section{Action of the algebra on Nekrasov partition functions}
\subsection{Gaiotto state}
The Drinfeld currents have a very remarkable action on Gaiotto states that can be written in terms of the operators $\CY^\pm$. The corresponding expressions, obtained after imposing a certain constraint on the Chern-Simons level, can be written in a projected form which removes singularities at infinity/origin. The projectors on the positive/negative powers in the expansion at $z=\infty$, will be denoted $\pr_\infty^\pm$, they are defined as follows:
\begin{equation}
\pr_\infty^++\pr_\infty^-=1,\quad \pr_\infty^+F(z)=\oint_\infty\dfrac{F(w)}{z-w}\dfrac{dw}{2i\pi},
\end{equation}
or explicitly for any function $F(z)$ expanded at $z=\infty$ as
\begin{equation}
F(z)=\sum_{k=-d}^\infty F_kz^{-k},\quad \pr_\infty^-F(z)=\sum_{k=1}^{\infty}F_kz^{-k},\quad \pr_\infty^+F(z)=\sum_{k=0}^d F_{-k}z^k.
\end{equation}
Similarly, $\pr_0^\pm$ denotes the projection on the positive/negative powers in the expansion at $z=0$:\footnote{The role of $\pr^\pm$ is naturally exchanged here since the singular part of the expansion contains (strictly) negative powers at $z=0$ and positive powers at $z=\infty$.}
\begin{align}
\begin{split}\label{def_pr0}
&\pr_0^++\pr_0^-=1,\quad \pr_0^-F(z)=\oint_0\dfrac{F(w)}{z-w}\dfrac{dw}{2i\pi},\\
&F(z)=\sum_{k=-d}^\infty F_kz^k,\implies \pr_0^-F(z)=\sum_{k=1}^{d}F_{-k}z^{-k},\quad \pr_0^+F(z)=\sum_{k=0}^\infty F_kz^k.
\end{split}
\end{align}

It is now possible to present one of the main results of this article: the actions of the Drinfeld currents on a Gaiotto state. We refer the reader to appendix \ref{AppB} for the details of the derivation. 
\begin{align}
\begin{split}\label{Gaiotto_right}
&e_+(z)\Ga_\kappa=- r\nu{\rm P}_\infty^-\left(z^{n-\kappa}\CY^+(zq_3^{-1})\right)\Ga_\kappa,\quad {\rm for}\ \kappa\leq 0,\\
&e_-(z) |G,\vec t\rangle_\kappa = r\mathrm{P}^+_0 (z^{-\kappa}\CY^-(z))|G,\vec t\rangle_\kappa\,,\quad {\rm for}\ \kappa> n, \\
&f_+(z)\Ga_\kappa= r{\rm P}_\infty^-\left(\frac{z^{\kappa-n}}{\CY^+(z)}\right)\Ga_\kappa,\quad {\rm for}\ \kappa\geq0,\\
&f_-(z)\Ga_\kappa=- r\nu q_3^{n}{\rm P}_0^+\left(\frac{z^{\kappa}}{\CY^-(zq_3)}\right)\Ga_\kappa,\quad {\rm for}\ \kappa< n ,
\end{split}
\end{align}
where we have introduced a shortcut notation for the normalization constant 
\begin{equation}\label{def_r}
r=\dfrac1{\sqrt{(1-q_1)(1-q_2)}}.
\end{equation}
We note that positive modes of the Drinfeld currents involve a projection at infinity to remove an unwanted pole contribution, while negative modes involve a projection at the origin. These compact expressions are only possible upon a restriction on the Chern-Simons level parameter. The dual actions on the bra Gaiotto state are given by the following expressions,
\begin{align}
\begin{split}\label{Gaiotto_left}
&_\kappa\bGa e_+(z)=r {\rm P}_\infty^-\left({}_\kappa\bGa\dfrac{z^{-\kappa}}{\CY^+(z)}\right),\quad {\rm for}\ \kappa\leq n\,\\
&_\kappa\bra{G,\vec{t}}e_-(z)=-r \nu q_3^n{\rm P}^+_0\left({}_\kappa\bGa\dfrac{z^{n-\kappa}}{\CY^-(zq_3)}\right),\quad {\rm for}\ \kappa > 0,\\
&_\kappa\bGa f_+(z)=-r\nu\pr_\infty^-\left({}_\kappa\bGa z^{\kappa}\CY^+(zq_3^{-1})\right),\quad {\rm for}\ \kappa\geq n,\\
&_\kappa\bGa f_-(z)=r\pr_0^+\left({}_\kappa\bGa z^{\kappa-n}\CY^-(z)\right),\quad {\rm for}\ \kappa<0.
\end{split}
\end{align}
Expanding these results either at $z=\infty$ or $z=0$ in the case of vanishing Chern-Simons level $\k=0$ produces the characterization of the Whittaker states in the rank $n$ representation,
\begin{align}
\begin{split}
&f_0\Ga=-\nu r q_3^n\Ga,\quad f_{k}\Ga=0,\quad k=1\cdots n-1,\quad f_{n}\Ga=r\Ga,
\end{split}
\end{align}
In the case $n=1$, these conditions reproduce the characterization of the Whittaker state for the Ding-Iohara algebra employed in \cite{feigin2011}

\subsection{Intertwiner}
When the representation spaces carry a Chern-Simons index, the interwiner operator defined in (\ref{def_intertwn}) has to be modified into
\begin{eqnarray}
V_{12}(\vec t_1,\vec t_2|\mu, \kappa, \kappa' ):\CV^{(\k)}\to\CV^{(\k')},\quad V_{12}(\vec t_1,\vec t_2|\mu, \kappa, \kappa' )=\sum_{\vec{Y}_1,\vec{Y}_2}\bZbf(\vec{t}_1,\vec{Y}_1;\vec{t}_2,\vec{Y}_2|\mu)\ket{\vec{t}_1,\vec{Y}_1}_{\kappa \;\kappa'}\bra{\vec{t}_2,\vec{Y}_2}.
\end{eqnarray}
The action of the positive/negative modes of the Drinfeld currents $e(z)$ and $f(z)$ on this deformed intertwiner is evaluated in appendix \ref{AppB}. It is observed that the action simplifies when the Chern-Simons level $\k$ and $\k'$ of the two representation spaces are related through 
\begin{equation}
\k'=\k+n_2-n_1,
\end{equation}
where $n_1$ and $n_2$ denote the rank of the representation. In this case, the action of the currents can be written in terms of the diagonal operators $\CY^\pm(z)$, suitably projected, and provided a proper behavior is assumed at $z=0$ or $z=\infty$. The latter condition imposes a further restriction on the range of the CS parameters.
\begin{align}
\begin{split}\label{comm_intertw}
&e_+(z) V_{12}(\mu,\k,\k')-\dfrac{\nu_1}{\nu_2}(q_3\mu)^{-\kappa'}\mu^{n_2} V_{12}(\mu,\k,\k')e_+(z\mu^{-1}q_3^{-1})=\\
&\hspace{3cm}- r\dfrac{\nu_1}{\nu_2}\mu^{n_2}\left[\pr_\infty^-\left(z^{-\kappa'}\CY^+(q_3^{-1}z) V_{12}(\mu,\k,\k')\dfrac1{\CY^+(q_3^{-1}\mu^{-1}z)}\right)\right]_+,\quad \k\leq 0,\\
&e_-(z) V_{12}(\mu,\k,\k')-\dfrac{\nu_1}{\nu_2}(q_3\mu)^{-\kappa'}\mu^{n_2} V_{12}(\mu,\k,\k')e_-(z\mu^{-1}q_3^{-1})=\\
&\hspace{3cm}r\left[\pr_0^+\left(z^{-\kappa}\CY^-(z) V_{12}(\mu,\k,\k')\dfrac1{\CY^-(\mu^{-1}z)}\right)\right]_-,\quad \k'>0,\\
&f_+(z) V_{12}(\mu,\k,\k')-\mu^{\k'-n_2}V_{12}(\mu, \kappa, \kappa')f_+ (z\mu^{-1})=\\
&\hspace{3cm}r\nu_2\mu^{-n_2}\left[\pr_\infty^-\left(\dfrac{z^{\kappa'}}{\CY^+(z)}V_{12}(\mu,\k,\k')\CY^+(zq_3^{-1}\mu^{-1})\right)\right]_+,\quad \k\geq 0,\\
&f_-(z) V_{12}(\mu,\k,\k')-\mu^{\k'-n_2}V_{12}(\mu, \kappa, \kappa')f_- (z\mu^{-1})=\\
&\hspace{3cm}-r\nu_1q_3^{n_1}\left[\pr_0^+\left(\dfrac{z^{\kappa}}{\CY^-(zq_3)}V_{12}(\mu,\k,\k')\CY^-(z\mu^{-1})\right)\right]_-,\quad \k'<0,
\end{split}
\end{align}
where we have employed the shortcut notation $V_{12}(\vec t_1,\vec t_2|\mu, \kappa, \kappa' )=V_{12}(\mu, \kappa, \kappa')$. It is easy to verify that when one of the representation space becomes trivial, the interwiner reduces to a Gaiotto state  (bra or ket), and the identities (\ref{Gaiotto_right}) and (\ref{Gaiotto_left}) are recovered.

\subsection{Flavor vertex operators}
The Chern-Simons and flavor vertex operators are special cases of a more general vertex operators depending on two sets of insertion points $\{z_i\}$ with $i\in I$ and $\{w_j\}$ with $j\in J$,
\ba
U_\pm(\{z_i\},\{w_j\})=\exp\lt(\sum_{i\in I}\Phi_\pm(z_i)-\sum_{j\in J}\Phi_\pm(w_j)\rt)\,.
\ea
The commutation relations with Drinfeld currents are derived by considering the action on the AFLT basis which form a faithful representation of the algebra,
\begin{align}
\begin{split}\label{com_vertex_ef}
&U_+(\{z_i\},\{w_j\})^{-1}e(u)U_+(\{z_i\},\{w_j\})=\frac{\prod_j(w_j-u)}{\prod_i(z_i-u)}e(u)\,,\\
&U_+(\{z_i\},\{w_j\})^{-1}f(u)U_+(\{z_i\},\{w_j\})=\frac{\prod_i(z_i-u)}{\prod_j(w_j-u)}f(u)\,,\\
&U_-(\{z_i\},\{w_j\})^{-1}e(u)U_-(\{z_i\},\{w_j\})=\frac{\prod_j(1-w_ju^{-1})}{\prod_i(1-z_iu^{-1})}e(u)\,,\\
&U_-(\{z_i\},\{w_j\})^{-1}f(u)U_-(\{z_i\},\{w_j\})=\frac{\prod_i(1-z_iu^{-1})}{\prod_j(1-w_ju^{-1})}f(u)\,.
\end{split}
\end{align}
Similar identities can be established for the positive/negative modes by applying the projections $\pr^-$ or $\pr^+$ to the Laurent expansion of the RHS. It is instructive to specialize to the case of the flavor vertex operator,
\begin{align}
\begin{split}\label{comm_vertex}
&\Uf(\vec m)^{\mp1}e(z)\Uf(\vec m)^{\pm 1}=\pf(zq_3^{-1})^{\mp 1}e(z),\quad \Uf(\vec m)^{\mp1}f(z)\Uf(\vec m)^{\pm1}=\pf(zq_3^{-1})^{\pm1} f(z),\\
&\Uaf(\vec\tm)^{\mp1}e(z)\Uaf(\vec\tm)^{\pm 1}=\paf(z)^{\mp 1}e(z),\quad \Uaf(\vec\tm)^{\mp1}f(z)\Uaf(\vec\tm)^{\pm1}=\paf(z)^{\pm1}f(z),
\end{split}
\end{align}
where we have introduced the mass polynomials in variables $z$ or $z^{-1}$ respectively,
\begin{equation}
\pf(z)=\prod_{f=1}^{n_f}(1-zm_f^{-1}),\quad \paf(z)=\prod_{f=1}^{\tn_f}(1-\tm_f z^{-1}).
\end{equation}

\section{qq-character and Ward identity}\label{s:Ward}
\subsection{Pure $U(n)$ gauge theories with a Chern-Simons term}
The simplest 5d $\CN=1$ SYM theory consists of a single $U(n)$ vector multiplet without any matter field. The Hilbert space contains only one copy of the rank $n$ representation space $\CV$. Including a Chern-Simons term, it is more convenient to consider two representation spaces $\CV^{(\k_L)}$ and $\CV^{(\k_R)}$ with the deformed scalar product (\ref{def_scalar}). A weighted trace for operators $\Op:\CV^{(\k_L)}\to\CV^{(\k_R)}$ acting in this space can be defined as the $\qf$-graded normalized expectation value in a Gaiotto state,
\begin{equation}\label{def_tr}
\la\Op\ra_{(\kappa_L, \kappa_R)}=\dfrac{{}_{\kappa_L}\bGa \qf^D\Op\Ga_{\kappa_R}}{{}_{\kappa_L}\bGa \qf^D\Ga_{\kappa_R}}=\dfrac1{{}_{\kappa_L}\bGa \qf^D\Ga_{\kappa_R}}\sum_{\vec Y}\qf^{|\vec Y|}\Zv(\vec t,\vec Y)\ _{\kappa_L}\baY \Op\aY_{\kappa_R}.
\end{equation}
The instanton partition function coincides with the normalization factor,
\begin{equation}
\Zinst=\ _{\k_L}\!\bGa \qf^D\Ga_{\k_R}.
\end{equation}
Two Chern-Simons levels have been introduced, $\k_L$ and $\k_R$, each associated to a different Hilbert space. However, due to the form of the scalar product (\ref{def_scalar}), the instanton partition function depends only on the difference of the Chern-Simons levels. Accordingly, all physical quantities are expected to depend only on the effective Chern-Simons level $\kappa_{eff}=\kappa_R-\kappa_L$. If so, the theory is invariant under the shift of $\kappa_L$ and $\kappa_R$ by an arbitrary integer, this extra symmetry will be fixed later.

The Ward identities can be deduced from the equivalence between the action of Drinfeld currents on the left and on the right for the Gaiotto states defining the trace. Considering the trace of $f_+(z)$, the two Chern-Simons levels must be restricted to $\kappa_L\geq n$ and $\kappa_R\geq 0$ to be able to employ the projection formulas given in (\ref{Gaiotto_right}), (\ref{Gaiotto_left}), and derive
\begin{equation}
\la f_+(z)\ra_{(\kappa_L, \kappa_R)}=r \pr_\infty^-\left(\la\frac {z^{\kappa_R-n}}{\CY^+(z)}\ra_{(\kappa_L, \kappa_R)}\right)=-r\qf^{-1}\pr_\infty^-\la\nu z^{\kappa_L}\CY^+(zq_3^{-1})\ra_{(\kappa_L, \kappa_R)},
\end{equation}
where the commutation relation with $\qf^D$ is easily obtained by noticing that $f(z)$ add boxes to $\baY$ when acting on the left. The Ward identity follows,
\ba\label{Ward}
\pr_\infty^-\left[z^{\kappa_L-n}\lt\langle \nu z^n\CY^+(zq_3^{-1})+\qf\frac {z^{\kappa_{eff}}}{\CY^+(z)}\rt\rangle_{(\kappa_L, \kappa_R)}\right]=0,
\ea
which corresponds to a particular linear combination of the discrete Ward identities presented in (\ref{discrete_Ward_Zv}). It suggests to define the qq-character as
\begin{equation}\label{def_qq1}
\chi_+(z)=\la\nu z^n\CY^+(zq_3^{-1})+\qf\dfrac{z^{\k_{eff}}}{\CY^+(z)}\ra_{(\kappa_L, \kappa_R)}.
\end{equation}
As seen from the formula (\ref{def_Y}), the eigenvalues of the operator $\CY^+(z)$ do not depend on the Chern-Simons level of the representation space. As a result, $\chi_+(z)$ only depends on the effective level $\k_{eff}$. Its asymptotic properties can be deduced from those of the eigenvalues of $\CY^+(z)$, and lead to the expansion
\begin{equation}
\chi_+(z)\simeq\sum_{k=-\infty}^d \chi_k^+ z^k,
\end{equation}
with $d=\text{max}(n,\keff)$. Setting $r=\k_L-n$, the Ward identity (\ref{Ward}) takes the form
\begin{equation}
\pr_\infty^-\left[\sum_{k=-\infty}^d \chi_k^+ z^{k+r}\right]=\sum_{k=-\infty}^{-r-1} \chi_k^+ z^{k+r}=0,
\end{equation}
or equivalently $\chi_k^+=0$ for $k<-r$. The strongest requirement is obtained for $r=0$ (or $\k_L=n$) leading to $\chi_k^+=0$ for $k<0$. It results that $\chi_+(z)$ is a polynomial of degree $\text{max}(n,\keff)$, and $\keff\geq -n$. Furthermore, the expression (\ref{Ward}) is reminiscent of the Seiberg-Witten curve $\CY+\qf/\CY\propto P_n(z)$, with $P_n(z)$ representing a polynomial of degree $n$ in $z$. In fact, it is shown in the next section that the qq-character degenerate to the polynomial $P_n(z)$ in the limit $q_1,q_2\to1$. It is thus natural to assume that it is of degree $d=n$. Given the previous constraints on the Chern-Simons level, the effective level of the theory is restricted to the range $-n\leq \kappa_{eff}\leq n$, in agreement with the bounds obtained in \cite{Intriligator-Morrison-Seiberg}.

The same result can also be derived by examination of the action of the positive modes of the Drinfeld current $e(z)$,
\begin{equation}
\la e_+(z)\ra_{(\kappa_L, \kappa_R)}=- r \pr_\infty^-\left[\nu z^{n-\kappa_R}\la\CY^+(zq_3^{-1})\ra_{(\kappa_L, \kappa_R)}\right]=\qf r \pr_\infty^-\left[z^{-\kappa_L}\la\dfrac1{\CY^+(z)}\ra_{(\kappa_L, \kappa_R)}\right],
\end{equation}
with the restrictions $\k_R\leq 0$, $\k_L\leq n$. The Ward identity reads 
\ba
\pr_\infty^-\left[z^{-\kappa_R}\lt\langle \nu z^n\CY^+(zq_3^{-1})+\qf\frac {z^{\kappa_{eff}}}{\CY^+(z)}\rt\rangle_{(\kappa_L, \kappa_R)}\right]=0.\nn
\ea
The strongest requirement is obtained by setting $\k_R=0$, and again imposes that $\chi_+(z)$ defined in (\ref{def_qq1}) is a polynomial, and $\keff\geq -n$. As explained previously, assuming that it is of degree exactly $n$ further constrains the effective level to be in the physical range $-n\leq \kappa_{eff}\leq n$.

A similar analysis can be performed for the negative modes of the Drinfeld currents, with the projections taken at the origin. For instance, the trace of $f_-(z)$ provides the Ward identity
\begin{equation}\label{Ward2}
\pr_0^+\left[z^{\k_L-n}\chi_-(z)\right]=0,\quad \text{with}\quad \chi_-(z)=\la\CY^-(z)+\qf\frac{z^{\kappa_{eff}+n}}{\nu q_3^n\CY^-(zq_3)}\ra_{(\kappa_L, \kappa_R)},
\end{equation}
for $\k_R<n$ and $\k_L<0$. Like $\chi_+(z)$, the qq-character $\chi_-(z)$ depends only on the difference $\keff$ of the two Chern-Simons levels. Introducing the asymptotic expansion of
\begin{equation}
z^{-n}\chi_-(z)\simeq \sum_{k=-d'}^\infty\chi_k^- z^k,\quad d'=-\text{min}(-n,\keff),
\end{equation}
into the Ward identity (\ref{Ward2}) gives
\begin{equation}
\pr_0^+\left[\sum_{k=-d'}^\infty\chi_k^- z^{k+\k_L}\right]=\sum_{k=-\k_L}^\infty\chi_k^-z^{k+\k_L}=0,
\end{equation}
equivalent to $\chi_k^-=0$ for $k\geq-\k_L$. This identity is valid for $\k_L<0$, and the strongest requirement is obtained after setting $\k_L=-1$ (so that $\keff=\k_R+1\leq n$). It provides $\chi_k^-=0$ for $k>0$, and
\begin{equation}
\chi_-(z)=\sum_{k=0}^{d'}\chi_{-k}^- z^{n-k}
\end{equation}
is a polynomial provided that $d'\leq n$ which is realized in the physical range $-n\leq\keff\leq n$. Due to the relation (\ref{rel_Y}) between the eigenvalues of the operators $\CY^+(z)$ and $\CY^-(z)$, the two qq-characters defined in (\ref{def_qq1}) and (\ref{Ward2}) describe the same quantity, i.e. $\chi_-(z)=\chi_+(z)$.\footnote{The qq-characters $\chi_+(z)$ (resp. $\chi_-(z)$) were primarily defined as an expansion around $z=\infty$ (resp. $z=0$). However, being polynomials, they can be analytically continued safely to the whole complex plane on which they coincide.} This is why in the following we will drop the index $\pm$ of the notation for the qq-character.

An explicit expression for $\chi(z)$ can easily be derived by expanding the RHS of (\ref{def_qq1}) at infinity, using the expression (\ref{def_Y}) for the eigenvalues. In the pure gauge case, $\keff=0$, and only the first term in (\ref{def_qq1}) contributes to the polynomial part,
\begin{equation}\label{expl_qq}
\chi(z)=\nu z^n\left(1+\dfrac1z\la\sum_{x\in R(\vec Y)}\chi_x-\sum_{x\in A(\vec Y)}q_3\chi_x\ra\rt)+{\cal O}(z^{n-2}).
\end{equation}
This result can also be expressed as a sum over the box content of $\vec Y$ using the shell formula (\ref{shell}), this is done in appendix \ref{AppE} up to the order $O(z^{n-3})$.

\subsection{qq-characters for super Yang-Mills with fundamental multiplets}\label{s:massive-qq}
The weighted trace associated to a $U(n)$ gauge theory with a number of fundamental and antifundamental flavors is normalized by the instanton partition function (\ref{Zinst_fund}). Its definition involves the introduction of flavor vertex operators in the expectation value of a Gaiotto state
\begin{align}
\begin{split}\label{def_tr_massive}
\la\Op\ra&=\dfrac{\bGa \qf^D \Uf(\vec m)\Uaf(\vec\tm)\Op\Ga}{\bGa \qf^D \Uf(\vec m)\Uaf(\vec\tm)\Ga}\\
&=\dfrac1{\Zinst}\sum_{\vec Y}\qf^{|\vec Y|}\Zv(\vec t,\vec Y)\Zf(\vec m;\vec t,\vec Y)\Zaf(\vec\tm;\vec t,\vec Y)\ \baY \Op\aY.
\end{split}
\end{align}
Here, for simplicity, we have turned off the Chern-Simons levels $\kappa_{L,R}$. 

In order to derive the regularity of the qq-character, we need to know the action of the positive/negative modes of the Drinfeld currents $e(z)$ and $f(z)$ on the flavor
vertex operators. Unfortunately, this action was only given in terms of the full currents in (\ref{comm_vertex}), and these relations need to be projected on the proper modes. For the sake of the argument, let us consider the positive modes $e_+(z)$ and the fundamental mass operator. The relation (\ref{comm_vertex}) involving the full current $e(z)$ reads
\begin{equation}
\Uf(\vec m)e(z)=\pf(zq_3^{-1}) e(z)\Uf(\vec m)
\end{equation}
where $\pf(z)$ is a polynomial of degree $n_f\leq n$. The action of $e_+(z)$ on a state $\aY$ is obtained from $e(z)$ by projecting out the positive powers of $z$ at infinity, so that formally $e_+(z)=\pr_\infty^-e(z)$. Using the property that for any polynomial $p(z)$ we have $\pr_\infty^-p(z)\pr_\infty^+=0$ and $\pr_\infty^-p(z)\pr_\infty^-=\pr_\infty^-p(z)$, it is shown that\footnote{This property is due to the fact that for any rational function $F(z)$, $\pr_\infty^+ F(z)$ is a polynomial, and so is $p(z)\pr_\infty^+ F(z)$. Expanded at $z=\infty$, a polynomial has no negative powers of $z$, and so $\pr_\infty^-p(z)\pr_\infty^+=0$. The second relation is deduced from the definition of the dual  projector $\pr_\infty^-=1-\pr_\infty^+$.}
\begin{equation}\label{comm_e_Uf}
\Uf(\vec m)e_+(z)=\pr_\infty^-\left[\pf(zq_3^{-1}) e(z)\Uf(\vec m)\right]=\pr_\infty^-\left[\pf(zq_3^{-1}) e_+(z)\Uf(\vec m)\right].
\end{equation}
A similar relation can be obtained for the antifundamental flavor vertex operator, exploiting the fact that $z^{\k}\paf(z)$ is a polynomial for $\k\geq\tnf$. For this purpose, it is useful to introduce the identity in the form $\CU^{-\k}\CU^\k=1$ in the trace (\ref{def_tr_massive}), and consider the commutation relation of 
\begin{equation}
\CU^\k\Uaf(\vec\tm)e(z)=z^\k\paf(z)e(z)\CU^\k\Uaf(\vec\tm),
\end{equation}
which can be projected on positive modes, 
\begin{align}
\begin{split}\label{comm_e_Uaf}
\CU^\k\Uaf(\vec\tm)e_+(z)&=\pr_\infty^-\left[\CU^\k\Uaf(\vec\tm)e(z)\right]=\pr_\infty^-\left[z^\k\paf(z)e(z)\CU^\k\Uaf(\vec\tm)\right]\\
&=\pr_\infty^-\left[z^\k\paf(z)e_+(z)\CU^\k\Uaf(\vec\tm)\right].
\end{split}
\end{align}

These properties can be applied to derive the regularity of the qq-character. First, consider the action on the right (\ref{Gaiotto_right}),
\begin{equation}
\la e_+(z)\ra=-r\nu\pr_\infty^-z^n\la\CY^+(zq_3^{-1})\ra,
\end{equation}
and then on the left, inserting the two spurious Chern-Simons levels, taking $\k\geq\tnf$:
\begin{align}
\begin{split}
\la e_+(z)\ra&=\dfrac1{\Zinst}\bGa \qf^D \Uf(\vec m)\Uaf(\vec\tm)\CU^{-\k}\CU^\k e_+(z)\Ga\\
&=\dfrac1{\Zinst}\pr_\infty^-\left[z^\k\pf(zq_3^{-1})\paf(z)\ _{\k\!}\bGa \qf^D  e_+(z)\Uf(\vec m)\Uaf(\vec\tm)\Ga_{\k}\right]\\
&=\qf r\pr_\infty^-\left[z^\k\pf(zq_3^{-1})\paf(z)\pr_\infty^-\la\dfrac{z^{-\k}}{\CY^+(z)}\ra\right]\\
&=\qf r\pr_\infty^-\la\dfrac{\pf(zq_3^{-1})\paf(z)}{\CY^+(z)}\ra.
\end{split}
\end{align}
The second equality is obtained using the commutation relations (\ref{comm_e_Uf}) and (\ref{comm_e_Uaf}), and then absorbing the Chern-Simons operators within the Gaiotto states. The third equality follows from the left action (\ref{Gaiotto_left}) on the Gaiotto state, which is obtained for $\k\leq n$ which implies $\tnf\leq n$. This inequality is always true for a physical theory. Finally, it is observed that the second projector can be omitted since $\k$ is assumed strictly positive in this computation. Combining the two previous results, we find that
\begin{equation}\label{qq_massive}
\pr_\infty^-\left[\chi^+(z)\right]=0,\quad \chi^+(z)=\la\nu z^n\CY^+(zq_3^{-1})+\qf\dfrac{m(z)}{\CY^+(z)}\ra,\quad m(z)=\pf(zq_3^{-1})\paf(z).
\end{equation}
which implies that $\chi(z)$ is a polynomial. For physical theories with $n_f+\tnf\leq n$, it is of degree $n$. Using the negative modes of the Drinfeld currents, it is possible to show that the qq-character defined around $z=0$ by
\begin{equation}
\chi^-(z)=\la\CY^-(z)+\qf\nu(zq_3)^n\dfrac{m(z)}{\CY^-(zq_3)}\ra
\end{equation}
is also a polynomial of degree $n$. In fact, due to the relation (\ref{rel_Y}) between the eigenvalues of operators $\CY^+(z)$ and $\CY^-(z)$, the two qq-characters are identical, namely $\chi^+(z)=\chi^-(z)$ for the analytic continuation to $z\in\mathbb{C}$.

It is instructive to compare the results obtained in this section with the work of Nekrasov, Pestun and Shatashvili \cite{NPS} (see also \cite{Kimura-Pestun}) where two operators were introduced,
\begin{equation}
\hat\chi_\text{NPS}^\pm(z)=\CY_\text{NPS}^\pm(z)+\dfrac{\CP^\pm(zq_3^{1/2})}{\CY_\text{NPS}^\pm(zq_3)}, \quad \CY^\pm(z)=\CY_\text{NPS}^\pm(zq_3^{\pm1/2}),
\end{equation}
with the following mass-dependent functions,
\begin{equation}
\CP^+(z)=\qf\nu^{-1}z^{-n} m(z),\quad \CP^-(z)=\qf \nu q_3^n z^n m(z).
\end{equation}
Using the previous identification, it is possible to relate our definition of the qq-characters to the operators defined in \cite{NPS},
\begin{align}
\begin{split}
&\hat\chi_\text{NPS}^-(q_3^{-1/2}z)=\CY^-(z)+\qf\nu(zq_3)^n\dfrac{m(z)}{\CY^-(zq_3)},\quad \hat\chi_\text{NPS}^+(zq_3^{-1/2})=\CY^+(zq_3^{-1})+\qf\nu^{-1}z^{-n}\dfrac{m(z)}{\CY^+(z)}\\
&\implies \chi^-(z)=\la\hat\chi_\text{NPS}^-(zq_3^{-1/2})\ra,\quad \chi^+(z)=\nu z^n\la\hat\chi_\text{NPS}^+(zq_3^{-1/2})\ra.
\end{split}
\end{align}

\subsection{Characters of linear quivers}
In a general quiver gauge theory, a representation space $\CV_i$ is attached to each node $i$ of the quiver. To each operator $\Op_i$ acting in the space $\CV_i$ can be associated the following weighted trace,
\begin{align}
\begin{split}
\la\Op_i\ra=\dfrac1{\Zinst}\sum_{\vec Y_k}&\prod_{\text{nodes }k}\qf_k^{|\vec Y_k|}\Zv(\vec t_k,\vec Y_k)\Zf(\vec m_k;\vec t_k,\vec Y_k)\Zaf(\vec\tm_k;\vec t_k,\vec Y_k)\ZCS(\k_k,\vec Y_k)\\
&\times\prod_{\text{links }<kl>}\Zbf(\vec t_k,\vec Y_k;\vec t_l,\vec Y_l|\mu_{kl})\times\ \bra{\vec t_i,\vec Y_i}\Op_i\ket{\vec t_i,\vec Y_i}.
\end{split}
\end{align}
This quantity can be obtained by insertion of the operator in the corresponding expectation value,
\begin{align}
\begin{split}\small
\la\Op_i\ra=\dfrac1{\Zinst}\langle G, \vec t_1|\qf_1^D\Uf(\vec m_1)\Uaf(\vec \tm_1)\UCS(\k_1)V_{12}(\vec t_1,\vec t_2|\mu_{12})\cdots\UCS(\k_i)\Op_iV_{ii+1}(\vec t_i,\vec t_{i+1}|\mu_{ii+1})\cdots|G, \vec t_\cdot\rangle.
\end{split}
\end{align}
A qq-character $\chi_i^{\pm}(z)$ is attached to each node of the quiver, and we can repeat the same procedure as in the one-node case to obtain their expressions, introducing a Drinfeld current $e_\pm(z)$ or $f_\pm(z)$ acting in the space $\CV_i$. Again, it is useful to deform the representation spaces by adding spurious Chern-Simons levels $\kappa_{Li}$ and $\kappa_{Ri}$ for the bra and ket states of the $i$-th node 
respectively, the final result depending only on the differences $\k_i=\k_{Ri}-\k_{Li}$.

As an illustration, the case of a quiver $A_2$ without fundamental/antifundamental matter fields is treated in appendix \ref{AppF}. There, it is assumed that the Chern-Simons levels belong to the physical range \cite{NPS},\footnote{It is also assumed that the theory is in the asymptotically free or conformal class, which restricts the ranks of the gauge groups to obey $\hf\leq\dfrac{n_1}{n_2}\leq 2$, thus ensuring that the physical range of Chern-Simons parameters is non-empty.}
\begin{equation}\label{phys_range}
-n_1\leq\k_1\leq n_1-n_2,\quad n_1-n_2\leq\k_2\leq n_2.
\end{equation}
Traces of operators are defined explicitly as,
\begin{align}
\begin{split}\label{def_traces_A2}
&\la\Op_1\ra=\dfrac1{\Zinst}\ _{\k_{L1}\!\!}\bra{G,\vec t_1}\qf_1^D\CO V_{12}(\vec t_1,\vec t_2|\mu,\k_{R1},\k_{L2})\qf_2^D\ket{G,\vec t_2}_{\k_{R2}},\\
&\la\Op_2\ra=\dfrac1{\Zinst}\ _{\k_{L1}\!\!}\bra{G,\vec t_1}\qf_1^DV_{12}(\vec t_1,\vec t_2|\mu,\k_{R1},\k_{L2})\qf_2^D\CO\ket{G,\vec t_2}_{\k_{R2}}.
\end{split}
\end{align}
Using the positive modes of the Drinfeld current, it is shown that the following qq-characters
\begin{align}
\begin{split}\label{def_chi12+}
\chi_1^+(z)&=\la\nu_1z^{n_1}\CY_1^+(zq_3^{-1})+\qf_1\nu_2\mu^{-n_2}z^{\k_1+n_2}\dfrac{\CY_2^+(zq_3^{-1}\mu^{-1})}{\CY_1^+(z)}+\qf_1\qf_2\mu^{-\k_2}\dfrac{z^{\k_1+\k_2}}{\CY_2^+(z\mu^{-1})}\ra,\\
\chi_2^+(z)&=\la\nu_2z^{n_2}\CY_2^+(zq_3^{-1})+\qf_2z^{\k_2}\dfrac{\CY_1^+(z\mu)}{\CY_2^+(z)}+\qf_1\qf_2\dfrac{\n_2}{\nu_1}q_3^{\k_1+n_2-n_1}\mu^{\k_1-n_1}\dfrac{z^{\k_1+\k_2+n_2-n_1}}{\CY_1^+(zq_3\mu)}\ra,
\end{split}
\end{align}
are polynomials in the physical range (\ref{phys_range}), of degree respectively $n_1$ and $n_2$. Similarly, it follows from the invariance under the action of the negative modes that the qq-characters
\begin{align}
\begin{split}\label{def_chi12-}
\chi_1^-(z)&=\la\CY_1^-(z)+\qf_1\nu_1q_3^{n_1}z^{\k_1+n_1}\dfrac{\CY_2^-(z\mu^{-1})}{\CY_1^-(zq_3)}+\qf_1\qf_2\nu_2q_3^{n_2}\mu^{-\k_2-n_2}\dfrac{z^{\k_1+\k_2+n_2}}{\CY_2^-(zq_3\mu^{-1})}\ra,\\
\chi_2^-(z)&=\la\CY_2^-(z)+\qf_2\dfrac{\nu_2}{\nu_1}q_3^{n_2-n_1}\mu^{-n_1}z^{\k_2+n_2-n_1}\dfrac{\CY_1^-(zq_3\mu)}{\CY_2^-(zq_3)}+\qf_1\qf_2\nu_2q_3^{\k_1+n_2+n_1}\mu^{\k_1}\dfrac{z^{\k_1+\k_2+n_2}}{\CY_1^-(zq_3^2\mu)}\ra,
\end{split}
\end{align}
are also polynomials of degree $n_1$ and $n_2$. It is observed that due to the relation (\ref{rel_Y}) between the eigenvalues of $\CY^+(z)$ and $\CY^-(z)$, the qq-characters defined in (\ref{def_chi12+}) and (\ref{def_chi12-}) are actually equivalent, namely $\chi_i^+(z)=\chi_i^-(z)$ for $i=1,2$.

The $\mathbb{Z}_2$-symmetry corresponding to exchanging the two nodes of the $A_2$ quiver involves a non-trivial mapping of the parameters that can be understood upon examining the bifundamental contribution,
\begin{equation}\label{sym_Zbf}
\Zbf(\vec t_1,\vec Y_1;\vec t_2,\vec Y_2|\mu)=\nu_1^{-|\vec Y_2|}\nu_2^{|\vec Y_1|}q_3^{-n_1|\vec Y_1|}\mu^{-n_1|\vec Y_2|-n_2|\vec Y_1|}\dfrac{\ZCS(n_2,\vec Y_1)}{\ZCS(n_1,\vec Y_2)}\Zbf(\vec t_2,\vec Y_2;\vec t_1,\vec Y_1|(\mu q_3)^{-1}).
\end{equation}
When the two nodes are exchanged, the rank of the two gauge groups are obviously mapped to each other, $n_1\leftrightarrow n_2$, and so are the central parameters $\vec t_1\leftrightarrow\vec t_2$ (and $\nu_1\leftrightarrow\nu_2$). On the other hand, due to the presence of extra factors in (\ref{sym_Zbf}), the Chern-Simons levels transform as
\begin{equation}
\k_1+n_2\to\k_2,\quad \k_2-n_1\to\k_1,
\end{equation}
the mass of the bifundamental multiplet is inverted and shifted, $\mu\to (q_3\mu)^{-1}$, and the gauge coupling parameters receive corrections in the form of extra factors:
\begin{equation}
\qf_1\nu_2\mu^{-n_2}\to\qf_2,\quad \qf_2\nu_1^{-1}(q_3\mu)^{-n_1}\to \qf_1.
\end{equation}
Under this symmetry, it is readily observed from (\ref{def_chi12+}) and (\ref{def_chi12-}) that the qq-character $\chi_1^\pm(z)$ is mapped to $\chi_2^\pm(z)$ and vice versa.

\section{Quantum Seiberg-Witten geometry}
In the limit $q_1,q_2\to 1$, the background reduces to $\mathbb{R}^4\times S_1$, and the expressions of the qq-characters $\chi^\pm$ in terms of the operators $\CY^\pm$ reproduce the Seiberg-Witten curve. In the case of $\CN=2$ 4d SYM, this fact is shown explicitly in \cite{BMZ}, built on previous works \cite{NPS,Bourgine2014a,Fucito2011,Poghossian2010,Nekrasov2003a}. A similar argument, although partly heuristic, can be given for the 5d theories we are studying here. For simplicity, we focus on the case of a single gauge group of rank $n$, with a Chern-Simons term of level $\k$, and fundamental/antifundamental matter fields,
\begin{equation}
\Zinst=\sum_{\vec Y}\qf^{|\vec Y|}\ZCS(\k,\vec Y)\Zf(\vec m;\vec t,\vec Y)\Zaf(\vec{\tilde{m}};\vec t,\vec Y)\Zv(\vec t,\vec Y).
\end{equation}

As a first step, we would like to derive the quantum Seiberg-Witten curve in the Nekrasov-Shatashvili limit $q_2\to1$. 
To perform this limit, we proceed by analogy with the 4d case and assume that the sum over $\vec Y$ is dominated by an $n$-tuple Young diagram $\vec Y^\ast$ that extremizes the sum. The critical Young diagrams composing $\vec Y^\ast$ are supposed to contain an infinite number of boxes, and obey the discrete saddle point equation
\begin{equation}
\dfrac{\qf^{|\vec Y^\ast+x|}\ZCS(\k,\vec Y^\ast+x)\Zf(\vec m;\vec t,\vec Y^\ast+x)\Zaf(\vec{\tilde{m}};\vec t,\vec Y^\ast+x)\Zv(\vec t,\vec Y^\ast+x)}{\qf^{|\vec Y^\ast|}\ZCS(\k,\vec Y^\ast)\Zf(\vec m;\vec t,\vec Y^\ast)\Zaf(\vec{\tilde{m}};\vec t,\vec Y^\ast)\Zv(\vec t,\vec Y^\ast)}=1
\end{equation}
for all $x\in A(\vec Y^\ast)$. Using the discrete Ward identities (\ref{discrete_Ward_Zv}), this condition can be written explicitly as
\begin{equation}\label{Bethe1}
\qf \dfrac{\chi_x^\k m(\chi_x)}{(1-q_1)(1-q_2)}\dfrac{\prod_{y\in R(\vec Y^\ast)}(1-\chi_x\chi_y^{-1})(1-\chi_y\chi_x^{-1}q_3^{-1})}{\prod_{\superp{y\in A(\vec Y^\ast)}{y\neq x}}(1-\chi_x\chi_y^{-1}q_3^{-1})(1-\chi_y\chi_x^{-1})}=1,
\end{equation}
with $m(z)=\pf(zq_3^{-1})\paf(z)$ a rational function containing the dependence in the mass of the fundamental/antifundamental fields. We further assume that $\vec Y^*$ is such that a box can be added or removed from each row $i$, so that
\begin{equation}
R(\vec Y^\ast)=\{(l,i,Y_i^{(l)}),\ i=1\cdots n_l,\ l=1\cdots n\},\quad A(\vec Y^\ast)=\{(l,i,Y_i^{(l)}+1),\ i=1\cdots n_l+1,\ l=1\cdots n\},
\end{equation}
where $n_l$ denotes the number of rows in the diagram $Y_l^\ast$, and $Y_i^{(l)}$ the number of boxes in each row. Under this condition, the equation (\ref{Bethe1}) can be expressed using the variables $e^{2\l_r}$ for the coordinates of the boxes in $R(\vec Y^\ast)$ with $r=1\cdots N_B=\sum_ln_l$, and $e^{2\xi_l}=t_l q_1^{n_l}$ for the $n$ extra boxes in $A(\vec Y^\ast)$,
\begin{equation}
\qf\dfrac{e^{2\k\l_r}m(e^{2\l_r})}{\prod_{l=1}^n (1-q_3^{-1}e^{2\l_r-2\xi_l})(1-e^{2\xi_l-2\l_r})}\prod_{\superp{s=1}{s\neq r}}^{N_B}\dfrac{(1-e^{2\l_r-2\l_s})(1-q_3^{-1}e^{2\l_s-2\l_r})}{(1-q_3^{-1}e^{2\l_r-2\l_s})(1-e^{2\l_s-2\l_r})}=1,
\end{equation}
where the factor $(1-q_1)(1-q_2)$ has been canceled by the term $(1-\chi_x\chi_y^{-1})(1-\chi_y\chi_x^{-1}q_3^{-1})$ for $\chi_{y\in R(\vec Y^\ast)}$ in the same column as $\chi_x$, i.e. $\chi_y=\chi_xq_2^{-1}$. It is now possible to set $q_2=1$, $q_3=q_1^{-1}=e^{-2i\a}$, and
\begin{equation}\label{Bethe2}
\tilde{\qf}e^{(2\k+n_f-\tnf)\l_r}\dfrac{\prod_{f=1}^{n_f}\sinh(\l_r-\mu_f+i\a)\prod_{\tilde{f}=1}^{\tnf}\sinh(\l_r-\tilde{\mu}_f)}{\prod_{l=1}^n\sinh(\l_r-\xi_l+i\a)\sinh(\l_r-\xi_l)}\prod_{\superp{s=1}{s\neq r}}^{N_B}\dfrac{\sinh(\l_r-\l_s-i\a)}{\sinh(\l_r-\l_s+i\a)}=1,
\end{equation}
where we have also introduced the notation $m_f=e^{2\mu_f}$, $\tilde{m}_f=e^{2\tilde{\mu}_f}$ for the mass parameters, and the renormalized gauge coupling constant,
\begin{equation}
\tilde{\qf}=\qf(-1)^{n+n_f}2^{n_f+\tnf-2n}e^{i\a (n_f-n)}e^{\sum_f\mu_f-\sum_{\tilde f}\tilde{\mu}_{\tilde{f}}}.
\end{equation}
The parameters $n_l$ (or $\xi_l$) play the role of cut-offs, so they must be sent to infinity at the end of the computation. The equations obtained in (\ref{Bethe2}) are reminiscent of the Bethe equations for the XXZ spin chain. In general, the integrable system pertaining to a 5d gauge theory compactified on a circle of radius $R$ is the relativistic (or anisotropic) version of the system obtained in the four dimensional case, and $1/R$ is identified with the speed of light (or anisotropy parameter) \cite{Nekrasov:1996cz}.

The limit $q_2\to1$ of the qq-character can be performed under the same assumptions, namely that the trace is dominated by the classical state $|\vec t,\vec Y^\ast\rangle$ with $\vec Y^\ast$ defined by the properties mentioned above. For this specific state, the eigenvalue of the operator $\CY^+(z)$ simplifies into
\begin{equation}
\tilde\CY^+(z,\vec Y^\ast)\simeq 2^ne^{\sum_l\xi_l-i\a N_B} e^{-nu}\prod_{l=1}^n\sinh(u-\xi_l)\prod_{r=1}^{N_B}\dfrac{\sinh(u-\l_r)}{\sinh(u-i\a-\l_r)},
\end{equation}
with the spectral parameter $z=e^{2u}$. Defining the Baxter $Q$-function
\begin{equation}
Q(u)=e^{-nu}\prod_{r=1}^{N_B}\sinh(u-\l_r),\quad \tilde\CY^+(z,\vec Y^\ast)\simeq 2^n(-1)^{n/2}\tilde{\nu}^{1/2} e^{-nu}\prod_{l=1}^n\sinh(u-\xi_l)\dfrac{Q(u)}{Q(u-i\a)},
\end{equation}
where $\tilde{\nu}=(-1)^ne^{2i\a N_B-2\sum_l\xi_l}$ is the limit of the parameter $\nu$.\footnote{Its expression can be obtained by noticing that the shell formula (\ref{shell}) specialized at $z=0$ implies
\begin{equation}
\nu=\dfrac{\prod_{x\in R(\vec Y)}(-\chi_x)}{\prod_{x\in A(\vec Y)}(-\chi_xq_3)}
\end{equation}}
Then, taking the limit $q_2\to1$ of the definition (\ref{qq_massive}), the qq-character can be written in the form of a Baxter TQ-equation associated to the set of Bethe equations (\ref{Bethe2}),
\begin{align}
\begin{split}\label{TQ}
T(u)Q(u)&=\prod_{l=1}^n\sinh(u+i\a-\xi_l) Q(u+i\a)\\
&+\tilde{\qf}e^{(2\k+n_f-\tnf)u}\ \dfrac{\prod_{f=1}^{n_f}\sinh(u+i\a-\mu_f)\prod_{f=1}^{\tnf}\sinh(u-\tilde{\mu}_f)}{\prod_{l=1}^n\sinh(u-\xi_l)}Q(u-i\a)
\end{split}
\end{align}
where $\chi^+(z)$ degenerates to the $T$-function,
\begin{equation}
T(u)=2^{-n}e^{-nu}(-1)^{n/2}\tilde{\nu}^{1/2}\lim_{q_2\to1}\chi^+(e^{2u}).
\end{equation}
In the case of pure gauge, $\k=n_f=\tnf=0$, we recover in the limit $\Re\xi_l\to\infty$ the TQ-equation of the relativistic periodic Toda chain with a twist,
\begin{equation}
\tilde{T}(u)\tilde{Q}(u)=\tilde{Q}(u+i\a)+\tilde{\qf}e^{2nu+ni\a}\tilde{Q}(u-i\a),
\end{equation}
with a minor modification of the functions $\tilde{T}(u)=e^{n(u+i\a)}T(u)$, $\tilde{Q}(u)=A^{-iu/\a}Q(u)$, $A=2^{-n}e^{\sum_l\xi_l}$.

In (\ref{TQ}), the cut-off dependent terms can be absorbed by a modification of the definition of the function $Q(u)$ which consists in introducing spurious poles,
\begin{equation}
Q(u)=\dfrac{e^{-nu}\prod_{r=1}^{N_B}\sinh(u-\l_r)}{\prod_{l=1}^n\prod_{k=1}^{n_l}\sinh(u-i\a(k-1)-\t_l)},
\end{equation}
with $t_l=e^{2\t_l}$ and $\xi_l=\t_l+i\a n_l$. Then, the TQ-equation no longer contains any cut-off variables,
\begin{align}
\begin{split}
T(u)Q(u)&=\prod_{l=1}^n\sinh(u+i\a-\t_l) Q(u+i\a)\\
&+\tilde{\qf}e^{(2\k+n_f-\tnf)u}\dfrac{\prod_{f=1}^{n_f}\sinh(u+i\a-\mu_f)\prod_{f=1}^{\tnf}\sinh(u-\tilde{\mu}_f)}{\prod_{l=1}^n\sinh(u-\t_l)}Q(u-i\a)
\end{split}
\end{align}
This TQ-equation can be written in the form of a quantum algebraic curve by introducing the shift operator
\begin{equation}
\hat y=e^{i\a\p_u},\quad [\hat y, u]=i\a\hat y,\quad \hat yz-q_1z\hat y=0,
\end{equation}
and it reads
\begin{align}
\begin{split}
T(u)Q(u)&=\left(\prod_{l=1}^n\sinh(u+i\a-\t_l)\hat y+\tilde{\qf}e^{(2\k+n_f-\tnf)u}\dfrac{\prod_{f=1}^{n_f}\sinh(u+i\a-\mu_f)\prod_{f=1}^{\tnf}\sinh(u-\tilde{\mu}_f)}{\prod_{l=1}^n\sinh(u-\t_l)}\hat y^{-1}\right)Q(u).
\end{split}
\end{align}
In this form, the limit $q_1\to1$ is easily obtained: the operator $\hat y$ becomes a variable $y$ commuting with $u$ (or $z$), and the function $Q(u)$ can be factorized out, leaving only
\begin{equation}
T(u)=y\prod_{l=1}^n\sinh(u-\t_l)+\tilde{\qf}y^{-1}e^{(2\k+n_f-\tnf)u}\dfrac{\prod_{f=1}^{n_f}\sinh(u-\mu_f)\prod_{f=1}^{\tnf}\sinh(u-\tilde{\mu}_f)}{\prod_{l=1}^n\sinh(u-\t_l)},
\end{equation}
or, in the original variables,
\begin{equation}
t(z)=y\prod_{l=1}^n(1-zt_l^{-1})+\qf(-1)^n y^{-1}\dfrac{z^{n+\k}m(z)}{\prod_{l=1}^n(1-z t_l^{-1})},\quad\text{ with}\quad t(z)=\lim_{q_1,q_2\to0}\chi^+(z).
\end{equation}
Here, $t(z)$ is a polynomial from its definition in terms of $\chi^+(z)$, and we recover the expression of the Seiberg-Witten curve for $\mathcal{N}=1$ SYM on $\mathbb{R}^4\times S_1$.\footnote{Defining further $\tilde{y}=yz^{-n/2}\prod_l(1-zt_l^{-1})$, it can be written in a more familiar form \cite{Nekrasov:1996cz},
\begin{equation}
\tilde{y}+\qf(-1)^n\dfrac{z^\k m(z)}{\tilde{y}}=z^{-n/2}t(z).
\end{equation}}

\section{Degenerate limit}
In the degenerate limit, the results presented in this article should reproduce those obtained in \cite{BMZ} and pertaining to the SH$^c$ algebra and 4d Nekrasov partition functions. To perform the limit procedure, it is necessary to express explicitly the dependence in the radius $R$ of the fifth dimension, which enters in the Kerov deformation parameters as $q_\a=e^{-R\e_a}$ where $\a=1,2,3$ and $\e_1+\e_2+\e_3=0$. 
It also appears in the central charge parameters $t_\ell=e^{-R a_\ell}$ when we try to associate the 4d Coulomb charges $a_\ell$ with $t_\ell$. 
We further denote $\chi_x=e^{-R\phi_x}$ with $\phi_x=a_\ell+(i-1)\e_1+(j-1)\e_2$ for a box $(\ell,i,j)\in\vec Y$, the spectral parameter $z=e^{-R\zeta}$, the mass parameter $\mu=e^{-Rm}$ and so on in terms of $R$. 
The radius $R$ is then sent to zero in order to recover the 4d background $\mathbb{R}^4_{\e_1,\e_2}$, and we consider its first non-trivial order.

As $R\to0$, it is easily seen from the expression (\ref{def_Y}) of their eigenvalues that the operators $\CY^\pm(z)$ degenerate to the operator $\CY_\text{BMZ}(z)$ introduced in \cite{BMZ},
\begin{equation}
\CY^+(z)\to(-R)^n\CY_\text{BMZ}(\z),\quad \CY^-(z)\to R^n\CY_\text{BMZ}(\z-\e_3),
\end{equation}
in agreement with the relation (\ref{rel_Y}), with $\nu\to(-1)^n$. As a result, the Cartan currents $\psi^\pm(z)$ which are expressed as a ratio of $\CY^\pm$ operators in (\ref{psi_YY}), degenerate into the Cartan generator of SH$^c$,
\begin{equation}\label{deg_psi}
\psi^\pm(z)\to R^2\e_1\e_2(-1)^n E(\z).
\end{equation}
On the other hand, the coefficients $\L_x(\vec Y)$ degenerate into the coefficients defined in \cite{BMZ} with the same notation up to a phase factor $(-1)^{n/2}$. As a consequence of the representation (\ref{repr_ef}), the positive/negative modes of the Drinfeld currents degenerate into the generating series of SH$^c$ generators with degree $\pm1$ as follows:
\begin{equation}
e_\pm(z)\to\mp R^{-1} (-1)^{n/2}D_{+1}(\z),\quad f_\pm(z)\to\mp R^{-1}(-1)^{n/2}D_{-1}(\z).
\end{equation}
The coefficient $\g_1$ behaves as $\g_1\sim R^3\e_1\e_2\e_3$, and the commutation relation (\ref{e-f}) reduces to one of the fundamental commutators defining the SH$^c$ algebra,
\begin{equation}
[e_\eta(z),f_\eta'(z')]\to\dfrac{(-1)^n}{R^2}[D_1(\z),D_{-1}(\z')]=\frac{(-1)^n}{R^2}\frac{\eta\eta'}{\e_3}\dfrac{E(\z)-E(\z')}{\z-\z'},
\end{equation}
where $\psi^+_0/\gamma_1$ drops out because it is of subleading order, $\psi_0^+$ being of order $O(R^2)$ as seen in (\ref{deg_psi}). Finally, the function $h(z)$ degenerates into a ratio of scattering factors for SH$^c$:
\begin{equation}
h(z)\to\dfrac{S(-\z)}{S(\z)},\quad S(\z)=\dfrac{(\z+\e_1)(\z+\e_2)}{\z(\z+\e_+)},
\end{equation} 
and the relations (\ref{e-e}) reproduce the braiding relations satisfied by the currents $D_{\pm1}(z)$ \cite{Tsymbaliuk}.

It is well-known that the degeneration of the 5d Nekrasov partition function reproduces the 4d partition function, with the same building blocks. Note however, that the Chern-Simons term $\ZCS(\k,\vec Y)\to 1+O(R)$ disappear in this limit. It is also possible to take the limit of the qq-character using the explicit expression provided in (\ref{expl_qq}) and recover the 4d qq-character for a theory with the same field content.

\section{Discussion}
In this article, we have analyzed the action of the quantum $\cW_{1+\infty}$ algebra $\cE$ on instanton partition functions of $\CN=1$ 5d SYM. 
For this purpose, the instanton partition functions have been built out of elements of the rank $n$ representation of the algebra: 
Gaiotto states, intertwiner, flavor and Chern-Simons vertex operators (or the Chern-Simons weighted Gaiotto state), 
encode the presence of respectively the gauge, bifundamental, fundamental/antifundamental multiplets and Chern-Simons terms. The action of the algebra on these building blocks has been given in the form of an action of the Drinfeld currents on the Gaiotto states (\ref{psi_YY},\ref{Gaiotto_right}-\ref{Gaiotto_left})  written in terms of the operators $\CY^\pm(z)$, and of their commutation relations with intertwiner and vertex operators (\ref{comm_intertw},\ref{com_vertex_ef}). These operators are elements of the Cartan of the algebra, diagonal in the AFLT basis, and can be used to define the qq-characters. The later encode, through its regularity property, a set of Ward identities (or recursion relations) obeyed by the instanton partition function. We have shown that this regularity follows from an invariance under the action of the Drinfeld currents $e(z)$ and $f(z)$. This derivation can easily be extended to arbitrary linear quivers, and an explicit expression has been provided for the qq-characters of the $A_2$ quiver. However, the treatment of $D,E$ quivers requires the introduction of a trivalent vertex whose transformation properties are still poorly understood. We hope to be able to provide more details on this issue in a future work.

Very recently, a deformed version of the algebra $\CE$ has attracted a lot of attention in the context of AGT correspondence and its extension to five and six dimensional gauge theories \cite{Awata:2011ce,Kimura-Pestun,Mironov:2016cyq,Mironov:2016yue}. It is obtained from $\CE$ by the introduction of a central element $\g$ that twist the commutation relations, and will be denoted here $\CE_\g$. As a result, the Cartan elements $\psi^\pm(z)$ of $\CE$ are no longer commuting in $\CE_\g$, and the representation on partition functions presented here no longer hold. In fact, the representation of the algebra $\CE_\g$ on the BPS sector of 5d $\CN=1$ SYM requires the introduction of a set of source terms coupled to the operators $D_k$ defined in (\ref{def_Dz}),
\begin{equation}
\Zinst(z)=\la \exp\left(t_0D_{0,1}-\sum_{n>0}(1-q_1^n)(1-q_2^n)t_n D_{0,n+1}\right)\ra.
\end{equation}
The Drinfeld currents generating $\CE_\g$ are then expressed in terms of the modes of a $q$-deformed Heisenberg algebra \cite{FHSSY,FHHSY,Kimura-Pestun,Mironov:2016yue} based on the bosonic modes $a_n=\p_{t_n}$ and $a_{-n}=t_n$. This underlying action of the algebra $\CE_\g$ account for both the Bethe/gauge and BPS/CFT correspondences. The correspondence with 2d conformal field theories is due to the presence of quantum versions of $q-$Virasoro and $\CW_n$ algebras in the rank $n$ representations \cite{FHSSY}. This is already an important feature of $\CE$ which can be verified in this paper. On the other hand, the deformation into $\CE_\g$ permits the construction of a TQ-system \cite{Kimura-Pestun}, thus rendering integrability at the level of the full Omega-background.\footnote{For instance, for gauge theories of rank $n=2$, the TQ-system consists in the stress-energy tensor of $q$-Virasoro and a $q$-analogue of the BRST charges obtained by $q$-integration of the screening!
  currents.} The qq-character studied here corresponds to the trace of the transfer matrix associated to the TQ-system. The algebras $\CE$ and $\CE_\g$ share a common subalgebra spanned by the positive modes of the Drinfeld currents $e_+(z)$, $f_+(z)$ and $\psi_+(z)$. It is clear from our derivation that the action of these modes only is sufficient to prove the regularity of the qq-character. This fact explains why the same property can be obtained in our simpler, untwisted, context. Furthermore, the operators $\CY^\pm(z)$ constructed here can be identified with the positive $q$-Heisenberg modes of similar operators defined in \cite{Kimura-Pestun} for $\CE_\g$. However, principally due to the fact that both $\CY^\pm(z)$ are diagonal, it does not seem possible to construct the full TQ-system in our algebraic framework. But it still provides a fast and efficient method for the derivation of the polynomial qq-characters encoding the Ward identities of the theory.

Finally we would like to mention the rapid developments on the connection between topological strings, the Ding-Iohara-Miki algebra
and instanton partition functions for 5d SYM \cite{Mironov:2016cyq, Mironov:2016yue, Awata:2016riz}.  In this approach, the Dynkin diagram of the gauge group and the quiver diagram are treated on an equal footing, and exchanged under S-duality. This treatment seems natural if we consider the whole algebra of symmetry $U_q(\mathfrak{gl}_r)$ affine for the $A_r$ quiver, whereas we focused in this paper on the diagonal subalgebra $\left(U_q(\mathfrak{gl}_1)\right)^{\otimes r}$. We hope to address the whole symmetry algebra in a future work, and construct algebraically the intertwiner operator representing the bifundamental contribution in this extended framework.

\section*{Acknowledgement}
We thank T. Kimura, K. Ohmori, A. Tsuchiya, J. Shiraishi for inspiring discussions. JEB would like to thank the IPhT CEA-Saclay for a warm hospitality, and especially S. Belliard, P. Di Francesco, R. Kedem, I. Kostov, D. Serban and V. Pasquier for discussions. JEB is supported by an I.N.F.N. post-doctoral fellowship within the grant GAST, and the UniTo-SanPaolo research  grant Nr TO-Call3-2012-0088 {\it ``Modern Applications of String Theory'' (MAST)}, the ESF Network {\it ``Holographic methods for strongly coupled systems'' (HoloGrav)} (09-RNP-092 (PESC)) and the MPNS--COST Action MP1210. YM is partially supported by Grants-in-Aid for Scientific Research (Kakenhi \#25400246) from MEXT, Japan.
HZ is supported by the National Research Foundation of Korea(NRF) grant 
funded by the Korea government(MSIP) (NRF-2014R1A2A2A01004951). 
RZ would like to thank sincerely the Hirose international scholarship for its generous financial support in the past seven years. 

\appendix
\section{Derivation of the identity (\ref{com_ef})}\label{AppA}
As an illustration of the derivation of the identity (\ref{com_ef}), we will only consider the commutation relation between the negative modes. It is obtained by projecting the commutator (\ref{e-f}) on positive powers of $z$ and $w$ in an expansion of the RHS,
\begin{align}
\begin{split}
\g_1[e_-(z),f_-(w)]&=\pr_{z=0}^+\pr_{w=0}^+\sum_{n\in\mathbb{Z}}\left(\dfrac{z}{w}\right)^n\sum_{k\geq 0}\left(\psi_k^+z^{-k}-\psi_{-k}^- z^k\right)\\
&=\psi_0^+-\sum_{k\geq 0}\psi_{-k}^-\sum_{n=0}^k w^n z^{k-n}.
\end{split}
\end{align}
The projection operator $\pr_0^+$ has been defined in (\ref{def_pr0}), here the variable on which it is acting has been made explicit. The modes $\psi_{k}^-$ are obtained from the current $\psi^-(z)$ after integration along a contour circling $z=0$,
\begin{equation}
\psi_{-k}^-=\dfrac1{2i\pi}\oint_0\dfrac{dx}{x} x^{-k}\psi^-(x).
\end{equation}
This formula is valid provided that the contour of integration does not include any of the current singularities. Inserting the integral in the expression of the commutator, we obtain
\begin{equation}
\g_1[e_-(z),f_-(w)]=\psi_0^+-\dfrac1{2i\pi}\oint_0\dfrac{dx}{x}\psi^-(x)\sum_{k\geq 0}\sum_{n=0}^k\left(\dfrac{w}{z}\right)^n\left(\dfrac{z}{x}\right)^k
\end{equation}
In order to take the sum, we assume $|w|<|z|$ and choose the contour of the integral such that $|z|<|x|$. Thus, this contour surrounds the origin but also the points at $x=z$ and $x=w$. Then we can use the formula
\begin{equation}
\sum_{k\geq 0}\sum_{n=0}^k\left(\dfrac{w}{z}\right)^n\left(\dfrac{z}{x}\right)^k=\dfrac{x^2}{(x-z)(x-w)},
\end{equation}
to rewrite the commutator as a contour integral that is easily evaluated by residues (under the assumption that $|z|$ is also smaller than the modulus of all the singularities in $\psi^-(x)$):
\begin{equation}
\g_1[e_-(z),f_-(w)]=\psi_0^+-\dfrac1{2i\pi}\oint_0dx\dfrac{x\psi^-(x)}{(x-z)(x-w)}=\psi_0^+-\dfrac{z\psi^-(z)-w\psi^-(w)}{z-w},\quad |w|<|z|.
\end{equation}
The other three commutators are evaluated using similar arguments.

\section{Check of the algebra for the rank $n$ representation}\label{a:algebra}
\subsection{Preliminaries}
Using the shell formula (\ref{shell}), the function $\PY(z)$ can also be written as a product over all the boxes in the Young diagrams,
\begin{equation}
\PY(z)=(1-q_1^{-1})(1-q_2^{-1})\prod_{\ell=1}^n\dfrac{1-zt_\ell^{-1}q_3^{-1}}{z-t_\ell}\prod_{x\in\vec Y}h(\chi_x/z).
\end{equation}
As a result, under a variation of the box content of Young diagrams, it is simply multiplied by a scattering factor,
\begin{equation}\label{var_Psi}
\dfrac{\Psi_{\vec Y\pm x}(z)}{\PY(z)}=h(\chi_x/z)^{\pm1}.
\end{equation}
Taking the residues, we deduce that for $x,y$ such that $\chi_x\neq q_\a^{\pm1}\chi_y$ we have the properties
\begin{equation}\label{var_L}
\left(\dfrac{\L_x(\vec Y\pm y)}{\L_x(\vec Y)}\right)^2=h(\chi_y/\chi_x)^{\pm 1},\quad\text{and}\quad \left(\dfrac{\L_x(\vec Y\pm x)}{\L_x(\vec Y)}\right)^2=1.
\end{equation}

\subsection{Commutation relations}
The commutation relations (\ref{p-e}) directly follows from the covariance of $\Psi_{\vec Y}(z)$:
\begin{align}
\begin{split}
\psi^{\pm}(z)e(w)\aY&=\sum_{x\in A(\vec Y)}\d(w/\chi_x)\L_x(\vec Y)\Psi_{\vec Y+x}(z)|\vec t,\vec Y+x\rangle\\
&=\sum_{x\in A(\vec Y)}\d(w/\chi_x)\L_x(\vec Y)h(\chi_x/z)\Psi_{\vec Y}(z)|\vec t,\vec Y+x\rangle\\
&=h(w/z)e(w)\psi^\pm(z),
\end{split}
\end{align}
and similarly for $f(z)$.

To check the commutation relation of $e(z)e(w)$, we assume $z\neq q_\a^{\pm1}w$ for $\a=1,2,3$, and consider
\begin{align}
\begin{split}
e(z)e(w)\ket{\vec{t},\vec{Y}}&=\sum_{x \in A(\vec{Y}),y\in A(\vec{Y}+x)}\delta(\chi_y/z)\delta(\chi_x/w)\Lambda_x(\vec{Y})
\Lambda_y(\vec{Y}+x)\ket{\vec{t},\vec{Y}+x+y}\\
&=\sum_{\substack{x \in A(\vec{Y}),y\in A(\vec{Y})\\x\neq y}}\delta(\chi_y/z)\delta(\chi_x/w)\Lambda_x(\vec{Y})
\Lambda_y(\vec{Y}+x)\ket{\vec{t},\vec{Y}+x+y}\\
&+\sum_{\substack{x \in A(\vec{Y}),y\in A(\vec{Y+x})\\y\notin A(\vec{Y})}}\delta(\chi_y/z)\delta(\chi_x/w)\Lambda_x(\vec{Y})
\Lambda_y(\vec{Y}+x)\ket{\vec{t},\vec{Y}+x+y}\,.
\end{split}
\end{align}
In the second term, we have either $\chi_y=q_1\chi_x$ or $\chi_y=q_2\chi_x$, which provides a delta function $\d(q_\a w/z)$ with $\a=1$ or $\a=2$. This delta function is never realized since $g(z,w)$ kills the non-vanishing contribution from $z= q_\a^{\pm1}w$, 
and it only remains
\begin{equation}
e(z)e(w)\ket{\vec{t},\vec{Y}}=\sum_{\substack{x \in A(\vec{Y}),y\in A(\vec{Y})\\x\neq y}}\delta(\chi_y/z)\delta(\chi_x/w)\Lambda_x(\vec{Y})
\Lambda_y(\vec{Y}+x)\ket{\vec{t},\vec{Y}+x+y}\nn\\
\end{equation}
Using
\begin{equation}
\L_x(\vec Y)\L_y(\vec Y+x)=h(\chi_x/\chi_y)\L_y(\vec Y)\L_x(\vec Y+y),
\end{equation}
we find
\begin{align}
\begin{split}
e(z)e(w)\ket{\vec{t},\vec{Y}}&=\sum_{\substack{x \in A(\vec{Y}),y\in A(\vec{Y})\\x\neq y}}\delta(\chi_y/z)\delta(\chi_x/w)h(\chi_x/\chi_y)\L_y(\vec Y)\L_x(\vec Y+y)\ket{\vec{t},\vec{Y}+x+y}\\
&=h(w/z) e(w)e(z)\aY.
\end{split}
\end{align}

Finally, we address the commutator between $e(z)$ and $f(w)$. It is expressed as a sum over states $|\vec t,\vec Y+x-y\rangle$ with either $x\in A(\vec Y)$ and $y\in R(\vec Y+x)$, or $y\in R(\vec Y)$ and $x\in A(\vec Y-y)$. Due to the properties (\ref{var_L}), only the diagonal terms $x=y$ remain, and using (\ref{Lx})
\begin{align}
\begin{split}
\lt[e(z),f(w)\rt]\ket{\vec{t},\vec{Y}}&=\lt(\sum_{x\in R(\vec{Y})}\chi_x^{-n}\delta(z/\chi_x)\delta(w/\chi_x)\Lambda_{x}^2(\vec{Y})
-\sum_{x\in A(\vec{Y})}\chi_x^{-n}\delta(z/\chi_x)\delta(w/\chi_x)\Lambda_{x}^2(\vec{Y})\rt)\ket{\vec{t},\vec{Y}}\nn\\
&=\dfrac1{\g_1}\delta(z/w)\lt(\sum_{x\in A(\vec{Y})}\delta(z/\chi_x)\chi_x^{-1}\Res_{w\to\chi_x}\Psi_{\vec Y}(w)+\sum_{x\in R(\vec{Y})}\delta(z/\chi_x)\chi_x^{-1}\Res_{w\to\chi_x}\Psi_{\vec Y}(w)\rt)\ket{\vec{t},\vec{Y}}\,.\label{comp-e-f}
\end{split}
\end{align}
The agreement with (\ref{e-f}) is obtained by decomposing the $\d$-function into positive and negative powers,
\begin{equation}
\chi_x^{-1}\d(z/\chi_x)=\left[\dfrac1{z-\chi_x}\right]_+-\left[\dfrac1{z-\chi_x}\right]_-
\end{equation}
where the first term should be expanded in powers of $z^{-1}$ as in $\psi^+(z)$, and the second one in powers of $z$ as in $\psi^-(z)$.

Eventually, to check the Serre relations, we perform a change of normalization to simplify the calculation. It is also possible to check it directly, however, the choice of the correct branch for the square roots in the expressions of $e$ and $f$ may be a little confusing. The renormalized basis is defined as 
\ba
\ket{\vec{t},\vec{Y}}'=\lt(\CZ_{\rm vect.}(\vec{t},\vec{Y})\rt)^{-1/2}\ket{\vec{t},\vec{Y}},\label{renorm}
\ea
and the representation for $e(z)$ takes the form 
\begin{equation}
e(z)\ket{\vec{t},\vec{Y}}'=r\sum_{x\in A(\vec{Y})}\delta(z/\chi_x)\L_x'(\vec Y)\ket{\vec{t},\vec{Y}+x}',\quad \L_x'(\vec Y)=\frac{\prod_{y\in R(\vec{Y})}1-\chi_y\chi_x^{-1}q_3^{-1}}{\prod_{\substack{y\in A(\vec{Y})\\y\neq x}}1-\chi_y\chi_x^{-1}}
\end{equation}
where the constant $r$ is expressed simply in terms of $q_1$ and $q_2$ and can be found in (\ref{def_r}). This representation is exactly the one presented in \cite{Tsymbaliuk} up to the overall factor $-q_3r^{-1}$, which is irrelevant for the consistency of the representation. 
Since $e(z)$ adds a box to the Young diagrams, the only non-vanishing matrix elements of the Serre relation for $e(z)$ are proportional to
\begin{equation}
^\prime\bra{\vec{t},\vec{Y}+x+y+z}[e_0,[e_1,e_{-1}]]\ket{\vec{t},\vec{Y}}'=r^3H(\chi_x,\chi_y,\chi_z;\vec{Y})\Lambda'_x(\vec{Y})\Lambda'_y(\vec{Y})\Lambda'_z(\vec{Y}),
\end{equation}
where we have introduced a renormalized bra vector orthonormal to the renormalized vectors. The function $ H(\chi_x,\chi_y,\chi_z;\vec{Y})$ can be expressed in terms of the rational function $\th(z)$ defined as
\begin{equation}
\th(z)=\dfrac{(1-z)(1-q_3^{-1}z)}{(1-q_1z)(1-q_2z)},\quad \dfrac{\L'_x(\vec Y+y)}{\L'_x(\vec Y)}=\th(\chi_y/\chi_x),
\end{equation}
as follows:
\begin{align}
\begin{split}
H(\chi_x,\chi_y,\chi_z;\vec{Y})&=\chi_x\chi_y^{-1}\lt(\theta(\chi_y/\chi_x)\theta(\chi_y/\chi_z)\theta(\chi_x,\chi_z)-\theta(\chi_x/\chi_y)\theta(\chi_x/\chi_z)\theta(\chi_y/\chi_z)\rt.\\
&\qquad\qquad\lt.-\theta(\chi_z/\chi_y)\theta(\chi_y/\chi_x)\theta(\chi_z/\chi_x)+\theta(\chi_z/\chi_x)\theta(\chi_z/\chi_y)\theta(\chi_x/\chi_y)\rt)\\
&+({\rm permutation\ among}\ x,\ y,\ {\rm and\ }z).
\end{split}
\end{align}
Here it is assumed that $x$, $y$ and $z$ are in a completely general position, i.e. any two of them do not sit side by side in the Young diagram. Contact terms must be treated separately, case by case. The terms we explicitly wrote down in the above expression come from the assignment that $x$ is added by $e_1$, $y$ is added by $e_{-1}$ and $z$ added by $e_0$. All other permutations cover the remaining assignments. 
After a tedious computation (or with Mathematica), it is seen that the coefficient $H(\chi_x,\chi_y,\chi_z;\vec{Y})$ are indeed vanishing. 
This implies that the rank $n$ representation also satisfies the Serre relation for $e(z)$ before the change of normalization. 

With the same redefinition of the ket vectors as in (\ref{renorm}), the action of $f(z)$ also reproduces the representation given in \cite{Tsymbaliuk} up to an overall factor $r^{-1}$:
\ba
f(z)\ket{\vec{t},\vec{Y}}'=rz^{-n}\sum_{x\in R(\vec{Y})}\delta(z/\chi_x)\frac{\prod_{y\in A(\vec{Y})}1-\chi_x\chi_y^{-1}q_3^{-1}}{\prod_{\superp{y\in R(\vec{Y})}{y\neq x}}1-\chi_x\chi_y^{-1}}\ket{\vec{t},\vec{Y}-x}'.
\ea
Serre relations are obtained using the same arguments as in the case of $e(z)$ treated previously.

\section{Recursion formula for the generalized Nekrasov factor}\label{a:recur-Nekra}

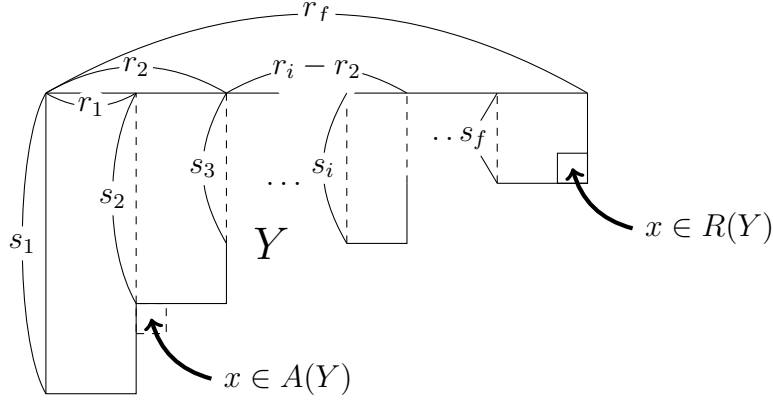
\begin{figure}[h]
\centering
\begin{tikzpicture}[scale=.4]
\coordinate (A)at(0,0);
\coordinate (B)at(0,-10);
\coordinate (C)at(3,-10);
\coordinate (D)at(3,-7);
\coordinate (E)at(6,-7);
\coordinate (F)at(6,-5);
\coordinate (G)at(18,-3);
\coordinate (H)at(18,0);
\coordinate (K)at(10,-5);
\coordinate (L)at(12,-5);
\coordinate (M)at(12,0);
\coordinate (P)at(12,-3);
\coordinate (N)at(10,0);
\coordinate (J)at(15,-3);
\coordinate (I)at(15,0);
\node at (7.5,-5) {\Large $Y$};
\node at (8,-3) {\dots};
\node at (13.5,-1.5) {\dots};
\draw (A)--(B)--(C)--(D)--(E)--(F);
\draw (H)--(A);
\draw (K)--(L)--(P);
\draw (J)--(G)--(H);
\draw[dashed] (D)--([yshift=7cm]D);
\draw[dashed] (F)--([yshift=5cm]F);
\draw[dashed] (K)--([yshift=5cm]K);
\draw[dashed] (P)--([yshift=3cm]P);
\draw[dashed] (J)--([yshift=3cm]J);
\draw[bend right, distance=2.1cm] (A) to node [fill=white, inner sep=0.2pt,circle] {$s_{1}$} (B);
\draw[bend right, distance=2.1cm] ([yshift=7cm]D) to node [fill=white, inner sep=0.2pt,circle] {$s_{2}$} (D);
\draw[bend right, distance=2.1cm] ([yshift=5cm]F) to node [fill=white, inner sep=0.2pt,circle] {$s_{3}$} (F);
\draw[bend right] (A) to node [fill=white, inner sep=0.2pt,circle] {$r_{1}$} ([yshift=7cm]D);
\draw[bend left] (A) to node [fill=white, inner sep=0.2pt,circle] {$r_{2}$} ([yshift=5cm]F);
\draw[bend left] (A) to node [fill=white, inner sep=0.2pt,circle] {$r_{f}$} (H);
\draw[bend left] ([yshift=5cm]F) to node [fill=white, inner sep=0.2pt,circle] {$r_{i}-r_2$} (M);
\draw[bend right, distance=2.1cm] (N) to node [fill=white, inner sep=0.2pt,circle] {$s_{i}$} (K);
\draw[bend right, distance=2.1cm] (I) to node [fill=white, inner sep=0.2pt,circle] {$s_{f}$} (J);
\draw[dashed] (D) rectangle ([xshift=1cm, yshift=-1cm]D);
\draw [] (G) rectangle ([xshift=-1cm, yshift=1cm]G);
\draw[->,bend left,ultra thick] ([xshift=2.5cm,yshift=-2.5cm]D)node[right]{$x\in A(Y)$}to([xshift=.5cm,yshift=-.5cm]D);
\draw[->,bend left,ultra thick] ([xshift=1.5cm,yshift=-1.5cm]G)node[right]{$x\in R(Y)$}to([xshift=-.5cm,yshift=.5cm]G);
\end{tikzpicture}
\caption{Rectangular decomposition of a Young diagram $Y$}\label{rectangular-decomp}
\end{figure}

In this appendix, we derive the recursion formula for the generalized Nekrasov factors, 
\ba
\tilde{N}_{Y,W}(t)=\prod_{(i,j)\in Y}F(t,W'_j-i,Y_i-j+1)
\prod_{(i,j)\in W}F(t,-Y'_j+i-1,-W_i+j)\,,
\ea
where $F(t,n,m)$ is an arbitrary function.
By setting it as
\ba
F(t,n,m)=\left\{
\begin{array}{ll}
	t+\e_1 m-\e_2 n\quad & \mbox{for 4d}\\
	1-tq_1^{-n}q_2^m & \mbox{for 5d}
	\end{array}
\right.,
\ea
we obtain the Nekrasov factor in 5d and 4d. The index $(i,j)$ refers to the box in the $i$th row and $j$th column. 
Consider the ratio $\frac{\tilde{N}_{Y+x,W}(t)}{\tilde{N}_{Y,W}(t)}$. 
By adding a box $x$ in the $I$-th row and $J$-th column of the Young diagram
$Y$, $Y_I$ and $Y'_J$ increase by $1$.
The ratio of the Nekrasov factor before and after the addition of the box becomes,
{\small \ba
\frac{\tilde{N}_{Y+x,W}(t)}{\tilde{N}_{Y,W}(t)}=
F(t,W'_J-I, Y_I-J+2) \prod_{j=1}^{Y_I}\frac{F(t,W'_j-I,Y_I-j+2)}{F(t,W'_j-I,Y_I-j+1)}
\prod_{i=1}^{W'_J}\frac{F(t,-Y'_J+i-2,-W_i+J)}{F(t,-Y'_J+i-1,-W_i+J)}
\ea}
In the second term, the cancellation between the numerator and the denominator 
occurs when $W'_j=W'_{j-1}$ and non-vanishing factors remain
at the corners of $W$  where $W'$ changes.  Similar cancellation
occurs in the third factor and again the non-vanishing factors lie
at the remaining corners of $W$.  In order to show the remaining terms
in the ratio, we introduce the rectangle decomposition of $Y$
and $W$ as shown in Fig. \ref{rectangular-decomp}.
We use the variables $r_i,s_i,\cdots$ to describe the shape of $Y$, and the bar variables $\bar{r}_i, \bar{s}_i,\cdots$ to encode the shape of $W$. After the cancellations, we find the formula 
\ba
\frac{\tilde{N}_{Y+x,W}(t)}{\tilde{N}_{Y,W}(t)}=
\frac{\prod_{p=1}^{\bar{f}+1} F(t, \bar{s}_p-I,J-\bar{r}_{p-1})}
{\prod_{p=1}^{\bar{f}}F(t, \bar{s}_p-I, J-\bar{r}_{p})}\,.
\ea

Similarly, we have 
\ba
\frac{\tilde{N}_{Y-x,W}(t)}{\tilde{N}_{Y,W}(t)}&=&\frac{\prod_{p=1}^{\bar{f}}F(t,\bar{s}_p-I,J-\bar{r}_{p})}
{\prod_{p=1}^{\bar{f}+1}F(t,\bar{s}_p-I,J-\bar{r}_{p-1})}\,,\\
\frac{\tilde{N}_{Y,W+x}(t)}{\tilde{N}_{Y,W}(t)}&=&\frac{\prod_{p=1}^{f+1}F(t,I-s_p-1,r_{p-1}+1-J)}
{\prod_{p=1}^{f}F(t,I-s_p-1,r_{p}-J+1)}\,,\nn\\
\frac{\tilde{N}_{Y,W-x}(t)}{\tilde{N}_{Y,W}(t)}&=&
\frac{\prod_{p=1}^{f}F(t,I-s_p-1,r_{p}-J+1)}{\prod_{p=1}^{f+1}F(t,I-s_p-1,r_{p-1}+1-J)}\,.
\ea
To recover the identities presented in (\ref{recursion-s})-(\ref{recursion-e}) is a matter of identification of the boxes coordinates in each set, for instance
\begin{equation}
A(Y)=\{(r_{p-1}+1,s_p+1)\text{ for }p=1\cdots f+1\},\quad R(Y)=\{(r_p,s_p)\text{ for }p=1\cdots f\}.
\end{equation}
 
\section{Action of the Drinfeld currents on the coherent states}\label{AppB}
\subsection{Gaiotto state}
The derivation of (\ref{Gaiotto_right}) and (\ref{Gaiotto_left}) follows what was done in \cite{BMZ} to treat the 4d case. Expanding the action of $e_\pm(z)$ on the $\aY_\k$ basis, we find
\begin{equation}
e_\pm(z)\Ga_\kappa=\mp \sum_{\vec Y}\sum_{x\in R(\vec Y)}\left[\dfrac{\L_x(\vec Y-x)\chi_x^{-\kappa}}{1-z\chi_x^{-1}}\right]_\pm \sqrt{\dfrac{\Zv(\vec t,\vec Y-x)}{\Zv(\vec t,\vec Y)}}\sqrt{\Zv(\vec t,\vec Y)}\aY_\kappa,
\end{equation}
which gives, using the recursion formulae,
\begin{align}
\begin{split}
e_\pm(z)\Ga_\kappa&=\mp r\sum_{\vec Y}\sum_{x\in R(\vec Y)}\left[\dfrac{\chi_x^{-\kappa}}{1-z\chi_x^{-1}}\right]_\pm\dfrac{\prod_{y\in A(\vec Y)}1-\chi_x\chi_y^{-1}q_3^{-1}}{\prod_{\superp{y\in R(\vec Y)}{y\neq x}}1-\chi_x\chi_y^{-1}}\sqrt{\Zv(\vec t,\vec Y)}\aY_\kappa\\
&=\mp r\sum_{\vec Y}\sum_{x\in R(\vec Y)}\left[\dfrac{1}{z-\chi_x}\right]_\pm \Res_{z\to\chi_x}z^{-\kappa}\frac{\prod_{y\in A(\vec{Y})}1-z\chi_y^{-1}q_3^{-1}}{\prod_{y\in R(\vec{Y})}1-z\chi_y^{-1}} \sqrt{\Zv(\vec t,\vec Y)}\aY_\kappa.
\end{split}\label{eGt}
\end{align}
To rewrite the summation over poles in a compact form, we use the following trick.
Let $g(z)$ be a rational function of the complex variable $z$.  We assume that
it has only simple poles, located at $z=z_i$ ($i=1,\cdots, N$), and possibly multiple poles at $z=0,\infty$.
Such a function can be written uniquely in the form,
\begin{eqnarray}
g(z)=g_+(z)+g_-(z) +\sum_{i=1}^N \frac{c_i}{z-z_i}, \quad g_+(z) = \sum_{n=0}^{N_+} g_{n} z^{n},\quad g_-(z)=\sum_{n=1}^{N_-} g_{-n} z^{-n},
\end{eqnarray}
where the coefficient $c_i$ are evaluated as residues of $g(z)$ at $z=z_i$,
$c_i=\Res_{z\to {z_i}}g(z)$ and  $g_\pm(z)$ are identified with $g_+ (z) = \mathrm{P}^+_\infty (g(z))$, $g_-(z) =\mathrm{P}^-_0(g(z))$.
One may apply the expansion at $z=\infty$ or $z=0$ by projecting out
the divergent part,
\begin{eqnarray}
\left[\mathrm{P}^-_\infty(g(z)) \right]_+ = g_-(z) +\sum_{i=1}^N \left[\frac{c_i}{z-z_i}\right]_+,\quad
\left[\mathrm{P}^+_0(g(z))\right]_- = g_+(z) +\sum_{i=1}^N \left[\frac{c_i}{z-z_i}\right]_-.
\end{eqnarray}
This trick can be applied to the RHS of (\ref{eGt}). In the case of $e_+(z)$, it can be expressed as a sum over the simple poles of the function $g(z)=z^{n-\kappa}\nu\tilde\CY^+(zq_3^{-1},\vec Y)$ at $z=\chi_x$ for $x\in R(\vec Y)$. The trick applies if the function $g(z)$ has no pole at $z=0$, so that $g_-(z)=0$. This condition imposes the restriction to $\kappa\leq 0$ which is deduced from the asymptotic form $\CY^+(z,\vec Y)\sim z^{-n}$ at $z=0$. In this range of the Chern-Simons level, the action of $e_+(z)$ can be expressed as
\ba
e_+(z) |G,\vec t\rangle_\kappa &=&-r \left[\sum_{\vec Y} \pr^-_\infty(z^{n-\kappa}\nu\tilde\CY^+(zq_3^{-1},\vec Y))
\sqrt{\Zv(\vec t,\vec Y)}\aY_\kappa\right]_+\nonumber\\
&=& -r\nu \mathrm{P}^-_\infty (z^{n-\kappa} \CY^+(zq_3^{-1}))|G,\vec t\rangle_\kappa\,.
\ea
A similar argument holds for the action of $e_-(z)$. In this case, the RHS of (\ref{eGt}) represents the sum over simple poles contribution of the function $g(z)=\tilde\CY^-(z,\vec Y)$ which is equivalent to the previous one due to the relation (\ref{rel_Y}). However, it is now required that the pole contribution at infinity vanishes, i.e. $g_+(z)=0$, which provides a different constraint over the Chern-Simons level, namely that $\kappa>n$. With this condition, (\ref{eGt}) becomes 
\ba
e_-(z) |G,\vec t\rangle_\kappa &=&r \left[\sum_{\vec Y} \pr^+_0(z^{-\kappa}\tilde\CY^-(z,\vec Y))
\sqrt{\Zv(\vec t,\vec Y)}\aY_\kappa\right]_-\nonumber\\
&=& r\mathrm{P}^+_0 (z^{-\kappa}\CY^-(z))|G,\vec t\rangle_\kappa\,.
\ea

The derivation of the left action on the bra is similar,
\ba
_\kappa\bra{G,\vec{t}}e_\pm(z)=\pm r\sum_{\vec{Y}}{}_\kappa\bra{\vec{t},\vec{Y}}\sqrt{\Zv(\vec t,\vec Y)}\sum_{x\in A(\vec{Y})}\left[\frac{1}{z-\chi_x}\right]_{\pm}\Res_{z\rightarrow \chi_x}z^{-\kappa}
\frac{\prod_{x\in R(\vec{Y})}1-\chi_xq_3^{-1}z^{-1}}{\prod_{x\in A(\vec{Y})}1-\chi_xz^{-1}}.
\ea
The right hand side can be written as the sum over the simple poles of $g(z)=z^{-\kappa}\tilde\CY^+(z, \vec{Y})^{-1}$ or $g(z)=\nu q_3^n z^{n-\k}\tilde\CY^-(zq_3,\vec Y)^{-1}$, leading to
\ba
_\kappa\bGa e_+(z)&=&r {\rm P}_\infty^-\left({}_\kappa\bGa\dfrac1{z^{\kappa}\CY^+(z)}\right),\quad {\rm for}\ \kappa\leq n\,,\\
_\kappa\bra{G,\vec{t}}e_-(z)&=&-r \nu q_3^n{\rm P}^+_0\left({}_\kappa\bGa\dfrac{z^{n-\kappa}}{\CY^-(zq_3)}\right),\quad {\rm for}\ \kappa> 0\,.
\ea

Similarly the action of the Drinfeld currents $f_\pm(z)$ decomposed on the $\aY$ basis reads
\begin{align}
\begin{split}
f_\pm(z)\Ga_\kappa&=\pm r\sum_{\vec Y}\sum_{x\in A(\vec Y)}\left[\dfrac{1}{z-\chi_x}\right]_{\pm}\Res_{z\to\chi_x}z^{\kappa-n}\frac{\prod_{x\in R(\vec{Y})}1-\chi_xq_3^{-1}z^{-1}}{\prod_{x\in A(\vec{Y})}1-\chi_xz^{-1}}\ \sqrt{\Zv(\vec t,\vec Y)}\aY_\kappa\,,\\
_\kappa\bra{G,\vec{t}}f_\pm(z)
&=\mp r\sum_{\vec{Y}}{}_\kappa\bra{\vec{t},\vec{Y}}\sqrt{\Zv(\vec t,\vec Y)}\sum_{x\in R(\vec{Y})}\left[\frac{1}{z-\chi_x}\right]_{\pm}\Res_{z\rightarrow \chi_x}z^{\kappa-n}
\frac{\prod_{x\in A(\vec{Y})}1-z\chi_x^{-1}q_3^{-1}}{\prod_{x\in R(\vec{Y})}1-z\chi_x^{-1}}.
\end{split}
\end{align}
Again, the calculation can be performed using the same rational functions with $\k$ replaced by $n-\k$ and the action on bra and kets exchanged,
\begin{align}
	\begin{split}
		f_+(z)\Ga_\kappa&= r{\rm P}_\infty^-\left(\frac{z^{\kappa-n}}{\CY^+(z)}\right)\Ga_\kappa,\quad {\rm for}\ \kappa\geq0,\\
		f_-(z)\Ga_\kappa&=- r\nu q_3^{n}{\rm P}_0^+\left(\frac{z^{\kappa}}{\CY^-(zq_3)}\right)\Ga_\kappa,\quad {\rm for}\ \kappa< n ,\\
		_\kappa\bGa f_+(z)&=-r\nu\pr_\infty^-\left({}_\kappa\bGa z^{\kappa}\CY^+(zq_3^{-1})\right),\quad {\rm for}\ \kappa\geq n,\\
		_\kappa\bGa f_-(z)&=r\pr_0^+\left({}_\kappa\bGa z^{\kappa-n}\CY^-(z)\right),\quad {\rm for}\ \kappa< 0.
\end{split}
\end{align}	

\subsection{Intertwiner}
The strategy to prove the two formulas (\ref{comm_intertw}) is same as in the case of Gaiotto states. We first analyze the action of $e_\pm(z)$ on the intertwiner. Using the discrete Ward identities on vector and bifundamental contributions, we find
\begin{align}
	\begin{split}
		e_\pm(z)V_{12}(\mu, \kappa, \kappa')= \pm r\sum_{\vec Y_1,\vec Y_2}\sum_{x\in R(\vec Y_1)}\chi^{1-\kappa}_x\left[\dfrac{1}{z-\chi_x}\right]_{\pm}\dfrac{\prod_{y\in A(\vec Y_1)}1-\chi_x\chi_y^{-1}q_3^{-1}}{\prod_{\superp{y\in R(\vec Y_1)}{y\neq x}}1-\chi_x\chi_y^{-1}}\dfrac{\prod_{y\in R(\vec Y_2)}1-\mu^{-1}\chi_x\chi_y^{-1}}{\prod_{y\in A(\vec Y_2)}1-\mu^{-1}\chi_x\chi_y^{-1}q_3^{-1}}\\
		\hspace{6cm}\times\bZbf(\vec{t}_1,\vec{Y}_1;\vec{t}_2,\vec{Y}_2|\mu)|\vec t_1,\vec Y_1\rangle_{\kappa \;\kappa'}\langle\vec t_2,\vec Y_2|\,,
\end{split}
\end{align}
and
\begin{align}
	\begin{split}
			(q_3\mu)^{-\kappa'}\mu^{n_2}\dfrac{\nu_1}{\nu_2}V_{12}(\mu, \kappa, \kappa')e_\pm (zq_3^{-1}\mu^{-1})= \pm r (q_3\mu)^{1-\kappa'}\mu^{n_2}\dfrac{\nu_1}{\nu_2}\sum_{\vec Y_1,\vec Y_2}\sum_{x\in A(\vec Y_2)}\chi^{1-\kappa'}_x\left[\dfrac{1}{z-\chi_xq_3\mu}\right]_{\pm}\\
			\hspace{-5.5cm}\times\dfrac{\prod_{y\in R(\vec Y_2)}1-\chi_x^{-1}\chi_yq_3^{-1}}{\prod_{\superp{y\in A(\vec Y_2)}{y\neq x}}1-\chi_x^{-1}\chi_y}\dfrac{\prod_{y\in A(\vec Y_1)}1-\mu^{-1}\chi_x^{-1}\chi_y}{\prod_{y\in R(\vec Y_1)}1-\mu^{-1}\chi_x^{-1}\chi_yq_3^{-1}}\bZbf(\vec{t}_1,\vec{Y}_1;\vec{t}_2,\vec{Y}_2|\mu)|\vec t_1,\vec Y_1\rangle_{\kappa \;\kappa'}\langle\vec t_2,\vec Y_2|\,.
	\end{split}
\end{align}
We focus on the case of $e_+(z)$ since the reasoning is the same for the negative modes. Then, these two expressions can be rewritten as a sum over residues of a rational function:
\begin{align}
\begin{split}
&e_+(z)V_{12}(\mu, \kappa, \kappa')=- r\dfrac{\nu_1}{\nu_2}\mu^{n_2}\sum_{\vec Y_1,\vec Y_2}\sum_{x\in R(\vec Y_1)}\left[\dfrac{1}{z-\chi_x}\right]_{+}\res_{z\to\chi_x}z^{n_1-n_2-\k}\dfrac{\tilde\CY^+(zq_3^{-1},\vec Y_1)}{\tilde\CY^+(z\mu^{-1}q_3^{-1},\vec Y_2)},\\
&\hspace{6cm}\times\bZbf(\vec{t}_1,\vec{Y}_1;\vec{t}_2,\vec{Y}_2|\mu)|\vec t_1,\vec Y_1\rangle_{\kappa \;\kappa'}\langle\vec t_2,\vec Y_2|\,\\
&(q_3\mu)^{-\kappa'}\mu^{n_2}\dfrac{\nu_1}{\nu_2}V_{12}(\mu, \kappa, \kappa')e_+ (zq_3^{-1}\mu^{-1})=r\dfrac{\nu_1}{\nu_2}\mu^{n_2}\sum_{\vec Y_1,\vec Y_2}\sum_{x\in A(\vec Y_2)}\left[\dfrac{1}{z-\chi_xq_3\mu}\right]_{+}\res_{z\to\chi_xq_3\mu}z^{-\k'}\dfrac{\tilde\CY^+(zq_3^{-1},\vec Y_1)}{\tilde\CY^+(z\mu^{-1}q_3^{-1},\vec Y_2)}\\
&\hspace{8cm}\times\bZbf(\vec{t}_1,\vec{Y}_1;\vec{t}_2,\vec{Y}_2|\mu)|\vec t_1,\vec Y_1\rangle_{\kappa \;\kappa'}\langle\vec t_2,\vec Y_2|\,\\
\end{split}
\end{align}
The two rational functions coincide when the Chern-Simons levels are related through $\kappa' = \kappa +n_2 -n_1$. In this case, it is possible to rewrite the action of $e_+(z)$ with the help of the diagonal operator $\CY^+(z)$, provided that no unwanted pole appears at the origin. This requirement of regularity at $z=0$ imposes an additional restriction to the range of the Chern-Simons parameters: we have to impose $\k\leq 0$ in order to obtain
\begin{align}
\begin{split}
e_+(z) V_{12}(\mu,\k,\k')-&\dfrac{\nu_1}{\nu_2}(q_3\mu)^{-\kappa'}\mu^{n_2} V_{12}(\mu,\k,\k')e_+(z\mu^{-1}q_3^{-1})=\\
&- r\dfrac{\nu_1}{\nu_2}\mu^{n_2}\left[\pr_\infty^-\left(z^{-\kappa'}\CY^+(q_3^{-1}z) V_{12}(\mu,\k,\k')\dfrac1{\CY^+(q_3^{-1}\mu^{-1}z)}\right)\right]_+.
\end{split}
\end{align}
A similar expression can be obtained for the action of the negative modes $e_-(z)$ in terms of the operators $\CY^-(z)$,
\begin{align}
\begin{split}
e_-(z) V_{12}(\mu,\k,\k')-&\dfrac{\nu_1}{\nu_2}(q_3\mu)^{-\kappa'}\mu^{n_2} V_{12}(\mu,\k,\k')e_-(z\mu^{-1}q_3^{-1})=\\
&r\left[\pr_0^+\left(z^{-\kappa}\CY^-(z) V_{12}(\mu,\k,\k')\dfrac1{\CY^-(\mu^{-1}z)}\right)\right]_-,
\end{split}
\end{align}
valid for $\k'=\k+n_2-n_1$ and $\k'>0$ (or $\k>n_1-n_2$) from the condition of regularity at infinity.

The demonstration for $f_\pm(z)$ is parallel to what has been done for the action of $e_\pm(z)$. In the case of $f_+(z)$, we find
\begin{align}
\begin{split}
&f_+(z)V_{12}(\mu, \kappa, \kappa')=r\nu_2\mu^{-n_2}\sum_{\vec Y_1,\vec Y_2}\sum_{x\in A(\vec Y_1)}\left[\dfrac{1}{z-\chi_x}\right]_{+}\res_{z\to\chi_x}z^{\k-n_1+n_2}\dfrac{\tilde\CY^+(zq_3^{-1}\mu^{-1},\vec Y_2)}{\tilde\CY^+(z,\vec Y_1)},\\
&\hspace{6cm}\times\bZbf(\vec{t}_1,\vec{Y}_1;\vec{t}_2,\vec{Y}_2|\mu)|\vec t_1,\vec Y_1\rangle_{\kappa \;\kappa'}\langle\vec t_2,\vec Y_2|\,\\
&\mu^{\k'-n_2}V_{12}(\mu, \kappa, \kappa')f_+ (z\mu^{-1})=-r\nu_2\mu^{-n_2}\sum_{\vec Y_1,\vec Y_2}\sum_{x\in R(\vec Y_2)}\left[\dfrac{1}{z-\chi_x\mu}\right]_{+}\res_{z\to\chi_x\mu}z^{\k'}\dfrac{\tilde\CY^+(zq_3^{-1}\mu^{-1},\vec Y_2)}{\tilde\CY^+(z,\vec Y_1)}\\
&\hspace{8cm}\times\bZbf(\vec{t}_1,\vec{Y}_1;\vec{t}_2,\vec{Y}_2|\mu)|\vec t_1,\vec Y_1\rangle_{\kappa \;\kappa'}\langle\vec t_2,\vec Y_2|\,\\
\end{split}
\end{align}
Again, the two rational functions involved coincide when $\k'=\k+n_2-n_1$. Imposing further $\k\geq 0$, the action of $f_+(z)$ can be written in terms of the operator $\CY^+(z)$ acting diagonally in each representation space:
\begin{equation}
f_+(z) V_{12}(\mu,\k,\k')-\mu^{\k'-n_2}V_{12}(\mu, \kappa, \kappa')f_+ (z\mu^{-1})=r\nu_2\mu^{-n_2}\left[\pr_\infty^-\left(\dfrac{z^{\kappa'}}{\CY^+(z)}V_{12}(\mu,\k,\k')\CY^+(zq_3^{-1}\mu^{-1})\right)\right]_+.
\end{equation}
Similarly, for $f_-(z)$ with $\k'=\k+n_2-n_1$ and $\k'<0$ we find
\begin{equation}
f_-(z) V_{12}(\mu,\k,\k')-\mu^{\k'-n_2}V_{12}(\mu, \kappa, \kappa')f_- (z\mu^{-1})=-r\nu_1q_3^{n_1}\left[\pr_0^+\left(\dfrac{z^{\kappa}}{\CY^-(zq_3)}V_{12}(\mu,\k,\k')\CY^-(z\mu^{-1})\right)\right]_-.
\end{equation}

\section{Concrete evaluation of the qq-character}\label{AppE}
In the pure gauge case, $\keff=0$, and the second term in (\ref{def_qq1}) does not contribute to the polynomial part of $\chi(z)$. Its only role is to cancel the poles coming from the first term. As a result, it is sufficient to expand the first term at $z=\infty$ since
\begin{equation}
\chi(z)=\pr_\infty^+\la\nu z^n\CY^+(zq_3^{-1})\ra,
\end{equation}
where the expectation value $\la\cdots\ra$ denotes the weighted trace (\ref{def_tr}) with Chern-Simons level turned off, i.e. $\k_L=\k_R=0$. The eigenvalues of the operator $\CY^+(z)$ can be expressed as a product of the box content of the $n$-tuple Young diagram $\vec Y$ upon using the shell formula (\ref{shell}),
\ba
\nu z^n\tilde\CY^+(zq_3^{-1},\vec Y)=\nu\prod_{l=1}^n(z-t_lq_3)\prod_{x\in\vec{Y}}\frac{(1-\chi_xq_1^{-1}/z)(1-\chi_xq_2^{-1}/z)}{(1-\chi_xq_3/z)(1-\chi_x/z)}.
\ea
Expanding the second product at $z=\infty$ leads to
\begin{align}
\begin{split}
\chi(z)=\nu\prod_{l=1}^n(z-t_lq_3)\times&\Bigg(1+(1-q_1)(1-q_2)\dfrac{q_3}{z}\la\sum_{x\in\vec Y}\chi_x\ra+\hf(1-q_1^2)(1-q_2^2)\dfrac{q_3^2}{z^2}\la\sum_{\superp{x,y\in\vec Y}{x\neq y}}\chi_x\chi_y\ra\\
&+(1-q_1)(1-q_2)(1+q_3^{-1})\dfrac{q_3^2}{z^2}\la\sum_{x\in\vec Y}\chi_x^2\ra+O(z^{-3})\Bigg).
\end{split}
\end{align}

\section{Regularity of the qq-characters: $A_2$ quiver case}\label{AppF}
In this appendix, we show the regularity of the qq-characters in the case of two nodes, with the Chern-Simons level $\k_1$ and $\k_2$, and no fundamental/antifundamental matter. We assume that the Chern-Simons levels belong to the physical range (\ref{phys_range}), and traces will be evaluated according to (\ref{def_traces_A2}).

We first analyze the trace of the modes $f_+(z)$. Using the right and left actions on Gaiotto states given in (\ref{Gaiotto_right}) and (\ref{Gaiotto_left}), it is shown that
\begin{align}
\begin{split}\label{fp12}
&\la f_{+,1}(z)\ra=-r\qf_1^{-1}\n_1\pr_\infty^-\left[z^{\k_{L1}}\la\CY_1^+(zq_3^{-1})\ra\right],\quad \k_{L1}\geq n_1,\\
&\la f_{+,2}(z)\ra=r\pr_\infty^-\left[z^{\k_{R2}-n_2}\la\dfrac1{\CY_2^+(z)}\ra\right],\quad \k_{R2}\geq0.
\end{split}
\end{align}
In addition, the commutation (\ref{comm_intertw}) with the intertwiner provides the identity
\begin{equation}\label{fp_comm}
\la f_{+,1}(z)\ra-\qf_2\mu^{\k_{L2}-n_2}\la f_{+,2}(z\mu^{-1})\ra=r\nu_2\mu^{-n_2}\pr_\infty^-\left[z^{\k_{L2}}\la\dfrac{\CY_2^+(zq_3^{-1}\mu^{-1})}{\CY_1^+(z)}\ra\right],\quad \k_{R1}\geq0
\end{equation}
for $\k_{L2}=\k_{R1}+n_2-n_1$. Replacing the LHS with the help of (\ref{fp12}) gives the condition
\begin{equation}
\pr_\infty^-\left[z^{\k_{L1}-n_1}\chi_1^+(z)\right]=0.
\end{equation}
From this point, the analysis is the same as what was done in the case of a single gauge group. These equations, valid for $\k_{L1}\geq n_1$, imply a tower of constraints, and the stronger requirement is obtained for $\k_{L1}=n_1$. We deduce that $\chi_1^+(z)$, which by definition only depends on the effective Chern-Simons levels $\k_i=\k_{Ri}-\k_{Li}$, is a polynomial of degree $n_1$. The other, spurious, Chern-Simons levels are then fixed to the values $\k_{R1}=\k_1+n_1$, $\k_{L2}=\k_1+n_2$ and $\k_{R2}=\k_1+\k_2+n_2$. The method is valid if $\k_{R2}\geq0$ and $\k_{R1}\geq0$, which corresponds to $\k_1\geq -n_1$ and $\k_1+\k_2\geq -n_2$. These conditions are always realized in the physical range (\ref{phys_range}).

A similar analysis can be performed for $e_+(z)$: using  (\ref{Gaiotto_right}), (\ref{Gaiotto_left}) and (\ref{comm_intertw}), we find for $\k_{L2}=\k_{R1}+n_2-n_1$,
\begin{align}
\begin{split}
&\la e_{+,1}(z)\ra=r\qf_1\pr_\infty^-\left[\la\dfrac{z^{-\k_{L1}}}{\CY_1^+(z)}\ra\right],\quad \k_{L1}\leq n_1,\\
&\la e_{+,2}(z)\ra=-r\nu_2\pr_\infty^-\left[z^{n_2-\k_{R2}}\la\CY_2^+(zq_3^{-1})\ra\right],\quad \k_{R2}\leq0,\\
&\la e_{+,2}(z)\ra-\qf_2\dfrac{\nu_2}{\nu_1}(q_3\mu)^{\k_{L2}}\mu^{-n_2}\la e_{+,1}(zq_3\mu)\ra=r\qf_2\pr_\infty^-\left[z^{-\k_{L2}}\la\dfrac{\CY_1^+(z\mu)}{\CY_2^+(z)}\ra\right],\quad \k_{R1}\leq0.
\end{split}
\end{align}
These three identities imply that
\begin{equation}
\pr_\infty^-\left[z^{-\k_{R2}}\chi_2^+(z)\right]=0,
\end{equation}
with $\chi_2^+(z)$ defined in (\ref{def_chi12+}). The strongest condition is obtained for $\k_{R2}=0$, it implies that $\chi_2^+(z)$ is a polynomial of degree $n_2$. Setting $\k_{R2}=0$, $\k_{L2}=-\k_2$, $\k_{R1}=n_1-n_2-\k_2$ and $\k_{L1}=n_1-n_2-\k_1-\k_2$, the conditions $\k_{L1}\leq n_1$ and $\k_{R1}\leq 0$ are equivalent to $\k_1+\k_2\geq -n_2$ and $\k_2\geq n_1-n_2$, they are always realized in the physical range (\ref{phys_range}).

The analysis of the negative modes is very similar. Starting from ($\k_{L2}=\k_{R1}+n_2-n_1$)
\begin{align}
\begin{split}
&\la f_{-,1}(z)\ra=r\qf_1^{-1}\pr_0^+\left[z^{\k_{L1}-n_1}\la\CY_1^-(z)\ra\right],\quad \k_{L1}<0,\\
&\la f_{-,2}(z)\ra=-r\nu_2q_3^{n_2}\pr_0^+\left[\la\dfrac{z^{\k_{R2}}}{\CY_2^-(zq_3)}\ra\right],\quad \k_{R2}<n_2,\\
&\la f_{-,1}(z)\ra-\qf_2\mu^{\k_{L2}-n_2}\la f_{-,2}(z\mu^{-1})\ra=-r\nu_1q_3^{n_1}\pr_0^+\left[z^{\k_{R1}}\la\dfrac{\CY_2^-(z\mu^{-1})}{\CY_1^-(zq_3)}\ra\right],\quad \k_{L2}<0,
\end{split}
\end{align}
we find 
\begin{equation}
\pr_0^+\left[z^{\k_{L1}-n_1}\chi_1^-(z)\right]=0,
\end{equation}
for all $\k_{L1}<0$. Choosing $\k_{L1}=-1$ leads to the strongest constraint, it imposes that $\chi_1^-(z)$ is a polynomial of degree $n_1$. Then, we have $\k_{R1}=\k_1+1$, $\k_{L2}=\k_1+1+n_2-n_1$ and $\k_{R2}=\k_1+\k_2+1+n_2-n_1$. The inequalities $\k_{R2}<n_2$ and $\k_{L2}<0$ are equivalent to $\k_1+\k_2\leq n_1$ and $\k_1\leq n_1-n_2$ which are always true in the physical range.

Finally, the action of $e_-(z)$ gives for $\k_{L2}=\k_{R1}+n_2-n_1$:
\begin{align}
\begin{split}
&\la e_{-,1}(z)\ra=-r\qf_1\nu_1q_3^{n_1}\pr_0^+\left[z^{n_1-\k_{L1}}\la\dfrac1{\CY_1^-(zq_3)}\ra\right],\quad \k_{L1}>0,\\
&\la e_{-,2}(z)\ra=r\pr_0^+\left[z^{-\k_{R2}}\la\CY_2^-(z)\ra\right],\quad \k_{R2}>n_2,\\
&\la e_{-,2}(z)\ra-\qf_2\dfrac{\nu_2}{\nu_1}(q_3\mu)^{\k_{L2}}\mu^{-n_2}\la e_{-,1}(zq_3\mu)\ra=-r\qf_2\dfrac{\nu_2}{\nu1}(q_3\mu)^{\k_{L2}-\k_{R1}}\mu^{-n_2}\pr_0^+\left[z^{-\k_{R1}}\la\dfrac{\CY_1^-(zq_3\mu)}{\CY_2^-(zq_3)}\ra\right],\quad \k_{L2}>0,
\end{split}
\end{align}
which implies for the qq-character defined in (\ref{def_chi12-}),
\begin{equation}
\pr_0^+\left[z^{-\k_{R2}}\chi_2^-(z)\right]=0,
\end{equation}
for all $\k_{R2}>n_2$. The strongest condition is obtained for $\k_{R2}=n_2+1$, it implies that $\chi_2^-(z)$ is a polynomial of degree $n_2$. The other levels are set to $\k_{L2}=n_2+1-\k_2$, $\k_{R1}=n_1+1-\k_2$, and $\k_{L1}=n_1+1-\k_1-\k_2$. The conditions $\k_{L1}>0$ and $\k_{L2}>0$ correspond to $\k_1+\k_2\leq n_1$ and $\k_2\leq n_2$, again always true in the physical range.


\providecommand{\href}[2]{#2}\begingroup\raggedright\endgroup

\end{document}